\documentclass[12pt,draftclsnofoot,onecolumn]{IEEEtran}

\ifCLASSINFOpdf
   \usepackage[pdftex]{graphicx}
\else
   \usepackage[dvips]{graphicx}
   \graphicspath{{../eps/}}
\fi

\usepackage[cmex10]{amsmath}
\usepackage{algorithmic}
\usepackage{array}
\usepackage{mdwmath}
\usepackage{mdwtab}
\usepackage{amsmath}
\usepackage{color}
\usepackage{cite}
\usepackage{stfloats}
\usepackage{subeqnarray}
\usepackage{graphicx}
\usepackage{subfigure}
\usepackage{amsmath}
\usepackage{amssymb}
\usepackage{amsthm}
\usepackage[ruled,vlined]{algorithm2e}

\begin{document}

\title{On Hybrid Pilot for Channel Estimation in Massive MIMO Uplink}

\author{Jiaming~Li,~\emph{Student Member~IEEE},~Chau~Yuen,~\emph{Senior Member~IEEE},\\~Dong~Li,~\emph{Member~IEEE},~Xianda~Wu,~\emph{Student Member~IEEE},\\~and~Han~Zhang,~\emph{Member~IEEE}

\thanks{H.~Zhang and J.~M.~Li are with the School of Physics \& Telecommunication Engineering, South China Normal University, Guangzhou 510026, China. (e-mail: zhanghan@scnu.edu.cn).}
\thanks{C.~Yuen is with Singapore University of Technology and Design, Singapore. (e-mail: yuenchau@sutd.edu.sg).}
\thanks{D.~Li is with the Faculty of Information Technology, Macau University of Science and Technology, Taipa, Macau, China. (e-mail: dli@must.edu.mo).}
\thanks{X.~D.~Wu is with the Department of Electrical and Computer Engineering, University of Macau, Macau, China. (e-mail: xianda.wu@connect.umac.mo).}
\thanks{H. Zhang is the corresponding author.}
}

\markboth{DRAFT ~\today}%
{Shell \MakeLowercase{\textit{et al.}}: Bare Demo of IEEEtran.cls for Journals}

\maketitle
\begin{abstract}
\vspace{-5mm}
This paper introduces a hybrid pilot-aided channel estimation technique for mitigating the effect of pilot contamination for the uplink of multi-cell multiuser massive MIMO systems. The proposed hybrid pilot is designed such that it enjoys the complementary advantages between time-multiplexed (TM) pilot and time-superimposed (TS) pilot, and thereby, allows superior solution to the conventional pilot schemes. We mathematically characterize the impact of hybrid pilot on the massive MIMO uplink by deriving a closed-form approximation for the uplink achievable rate. In large-number-of-antennas regime, we obtain the asymptotically optimal solution for hybrid pilot by jointly designing the TM pilot and the TS pilot. It is shown that either TM pilot or TS pilot has the advantages for large frame-size and limited frame-size transmission, respectively, while the hybrid pilot scheme can offer a superior performance to that employing either TM pilot or TS pilot. Numerical results demonstrate the effectiveness of the proposed design.
\end{abstract}


\IEEEpeerreviewmaketitle
\newpage

\section{Introduction}
\IEEEPARstart{M}{assive} multiple-input multiple-out (MIMO) technique at the base station (BS), was firstly proposed in \cite{Mazetta10}, and now has attracted tremendous interest in both academia and industry. Massive MIMO (also, known as large scale MIMO) has been widely recognized as a potential candidate for the key technologies of the future wireless communication systems \cite{5G1,5G2,5G3}.

Compared with the conventional MIMO technique, massive MIMO with time-division duplex (TDD) exhibits several remarkable features. First, by taking advantage of channel reciprocity, additional antennas significantly increase the spectral efficiency through spatial multiplexing \cite{SM1,SM2}. Second, large antenna arrays enables energy efficiency in both uplink (UL) and downlink (DL) transmission through coherent combining, and hence, provide an potential for cell-size shrinking \cite{HET13}. Third, when the number of BS antennas $M$ is sufficiently large, the simplest coherent combiner and linear precoder, e.g. the matched filter (MF), turn out to be optimal \cite{precoder1,precoder2}. Although promising, the ultimate performance of TDD massive MIMO is limited by the effect of pilot contamination, an unavoidable interference caused by the reuse of pilots (or nonorthogonality of pilots) among several adjacent cells, even for the asymptotic case $M \to \infty$.

In an effort to solve the problem of pilot contamination when performing UL channel estimation, several sophisticated pilot-aided schemes have been proposed. Typically, pilots are time-multiplexed with the data during the training phase, and henceforth are referred to as time-multiplexed (TM) pilots. Relying on the coordination between neighboring cells, the second-order statistical information about the user channels of neighboring cells is involved for channel estimation \cite{YHF13}. Based on the singular value decomposition (SVD), blind channel estimation scheme is proposed in \cite{SVD1,SVD2}, which is shown to be effective in mitigating pilot contamination. For a fixed size of training, a pilot-reuse scheme is provided in \cite{PR12}, aiming to maximize the UL achievable rate. In \cite{DA14}, a data-aided scheme is presented by employing the decision feedback information of data symbols to aid the channel estimation. In \cite{PD1,PD2,PD3}, the optimal designs for TM pilots by maximizing the sum spectral efficiency are proposed and discussed, and the authors in \cite{TS1} propose using downlink training with pilot contamination precoding to eliminate the effect of contamination. All these studies employing TM pilots lead to a similar conclusion that the data rate will decrease with increasing pilot-size, making system throughput limited, especially for the mobility case, where the channel coherent time is limited.

As an alternative to TM pilots, time-superimposed (TS) pilots have been studied in the context of channel acquisition in massive MIMO systems \cite{SP1,ACCESS15}. In comparison with TM pilots \cite{YHF13,SVD1,SVD2,PR12,DA14,PD1,PD2,PD3,TS1}, TS pilots require no additional time resource reserved for pilots, and thereby, can achieve a higher spectral efficiency \cite{SCI13}. More recently, the analysis of TS pilots in massive MIMO systems \cite{TVT16} illustrates its superiority for mitigating pilot contamination. However, the mixed type of pilots suffers from co-interference from data symbols, which generally limits its performance, especially in low signal-to-noise ratio (SNR) scenarios\cite{TCOM_SP10,TVT_SP09}.

In this paper, we take a further step than the previous literatures \cite{SP1,TVT16,SCI13,ACCESS15,TCOM_SP10,TVT_SP09}, and propose a new pilot-based scheme as an alternative to the conventional pilot-aided ones for mitigating pilot contamination in massive MIMO systems. To be specific, the pilots for the UL channel estimation comprise both TM pilots and TS pilots, henceforth can be referred to as hybrid pilots. The motivation behind the proposed design is twofold.
\begin{itemize}
  \item TM pilots with the aid of TS pilots can improve the estimation quality, while preserving transmission efficiency\cite{SP1}.
  \item TS pilots benefit from TM pilots by reducing the correlation between pilots and data \cite{SP1,TVT16}, and hence can provide substantial improvement of system performance.
\end{itemize}
Intuitively, hybrid pilot enjoys the advantages of both TM pilot and TS pilot, and thereby, is more flexible and robust to different transmission of practical relevance. The hybrid design of pilots, to the best of authors' knowledge, has not been addressed for multi-cell MIMO systems. To evaluate the proposed design, we mathematically characterize the impact of hybrid pilot on the performance of massive MIMO uplink, and demonstrate its effectiveness by deriving a closed-form approximation on signal-to-interference-plus-noise ratio (SINR) as well as cell throughput. In large-number-of-BS-antennas regime, we obtain the asymptotically optimal solutions for the hybrid design of pilots. Our result demonstrates that the time allocation between TM pilots and TS pilots, as well as power ratio between pilots and data, determine the UL spectral efficiency. Qualitative analysis and simulations show that the conventional TM pilot or TS pilot is effective for either large-frame-size or limited frame-size transmission, while the proposed hybrid pilot design can offer a superior solution than that employing conventional pilots \cite{PD2,TVT16}.

The rest of this paper is organized as follows: In Section II, we firstly describe the uplink multi-cell massive MIMO system model. In Section III, we introduce the hybrid pilots-based channel estimation technique, and then provide the analytical results for the UL achievable rate. In Section IV, we provide an iterative data-aided solution to improve the system performance. Along with the theoretical analysis, the asymptotically optimal solutions are given in Section V, which can explain the trends observed in Section VI simulations results. Finally, Section VII summarizes the main results and insights obtained in the paper.

\emph{Notations}: Boldface lower and upper case symbols represent vectors and matrices, respectively. The transpose, complex conjugate, and Hermitian transpose operations are denoted by $()^T$, $()^*$, and $()^H$, respectively. $\|\cdot\|$ denotes the Euclidian norm and $E[\cdot]$ is the statistical expectation. We use $\mathcal{CN}(a, b)$ to denote the circular symmetric complex Gaussian distribution with mean $a$ and covariance $b$. $\mathcal{O}(\cdot)$ denotes the big-O notation. $\to$ denotes the convergence as $M \to \infty$.

\section{System Model and Problem Formulation}

\subsection{Multi-cell Massive MIMO Uplink}
Consider a cellular network composed of $L$ hexagonal cells, each consisting a central $M$-antenna BS and $K$ ($K \leq M$) single-antenna user terminals (UTs) that share the same bandwidth. We focus on the uplink transmission without any kind of BS cooperation. The propagation channel coefficient between $m$th BS antenna of the $j$th cell and the $k$-th UT of the target cell, i.e., cell $1$, is $h_{j,k,m} = \sqrt{\beta_{j,k}} g_{j,k,m}$, where $\{\beta_{j,k}\}$ and $\{g_{j,k,m}\}$ are large scale fading and small scale fading, respectively. Specifically, $\{\beta_{j,k}\}$ model path-loss and shadowing that change slowly and thus can be assumed to be known at receiver, while $\{g_{j,k,m}\} \sim \mathcal{CN}(0,1)$ are identically independent distributed (i.i.d.) unknown random variables. Moreover, $\{h_{j,k,m}\}$ are assumed to be constant for the duration of $T$ symbols in time, where $T$ is the channel coherence time that limited by the mobility of users.

\subsection{Effect of Pilot Contamination}
Denote $\textbf{u}_{1}(t) = [u_{1,1}(t),\cdots, u_{1,M}(t)]^T$ as the received signal vector at over $M$ antennas at the BS of target cell, i.e. cell $1$ at time instant $t$
\begin{align}
\label{rxvector}
\textbf{u}_{1}(t) = \sum_{j=1}^{L}\sum_{k=1}^{K} \textbf{h}_{j,k} \widetilde{x}_{j,k}(t) + \textbf{n}_1(t)
\end{align}
where $\textbf{h}_{j,k} = [h_{j,k,1},\cdots,h_{j,k,M}]^T$, $\textbf{n}_1(t) = [n_{1,1}(t),\cdots,n_{1,M}(t)]^T$ with $n_{1,m}(t) \sim \mathcal{CN}(0,\sigma_n^2)$ being the additive white Gaussian noise (AWGN), and $\widetilde{x}_{j,k}(t)$ denotes the transmitted signal from $k$-th user at $j$ cell with unit power at time $t$.

In each frame of transmission, $\tau$ TM pilots are employed as the training overhead, given in the form $\textbf{x}_{j,k} = [x_{j,k}(1),\cdots,x_{j,k}(\tau)]$ $\in \mathbb{C}^{\tau}$. Then, we rewrite (\ref{rxvector}) as a matrix form $\textbf{U}_{1} = [\textbf{u}_{1}(1),\cdots,\textbf{u}_{1}(\tau)]$ $\in \mathbb{C}^{M \times \tau}$, which is given by
\vspace{-2mm}
\begin{align}
\label{U}
\textbf{U}_{1} = \sum_{k=1}^{K}\textbf{h}_{1,k}\textbf{x}_{1,k} + \sum_{j \neq 1}^{L}\sum_{k=1}^{K}\textbf{h}_{j,k}\textbf{x}_{j,k} + \textbf{N}_{1}.
\vspace{-3mm}
\end{align}
The least-squares (LS) estimate on channel vector $\textbf{h}_{1,k}$ of the $k$-th UT in the target cell can be obtained as \cite{SMKay}
\vspace{-3mm}
\begin{align}
\label{LS_contamination}
\hat{\textbf{h}}_{1,k} = \textbf{h}_{1,k} + \frac{1}{\tau} \sum_{j \neq 1}^{L}\sum_{k'=1}^{K}\textbf{h}_{j,k} \big(\textbf{x}_{j,k} \textbf{x}_{1,k}^H\big) + \frac{1}{\tau} \textbf{N}_{1}\textbf{x}_{1,k}^H.
\vspace{-3mm}
\end{align}
The above formula implies that the estimation on $\textbf{h}_{1,k}$ are contaminated by the channel vectors of other cells, unless each user to be assigned a unique orthogonal pilot, i.e., $\frac{1}{\tau}\textbf{x}_{j,k}\textbf{x}_{1,k}^H = 0$ if $j \neq 1$. In practical TDD mode, $T$ is limited by the mobility of users, therefore it is hard to ensure the orthogonality of pilot sequences in the multi-cell scenario as the number of overall users becomes large. Although the pilot-based schemes in \cite{PR12,Larsson16,Jinshi16} are proposed to improve the estimation quality in (\ref{LS_contamination}), the correlated pilot sequences in different cells, known as pilot contamination, causes capacity-limiting inter-cell interference even when $M \to \infty$.
\vspace{-1mm}
\section{Hybrid Pilot-aided UL Channel Estimation}
In this section, we study a hybrid pilot-aided channel estimation scheme, where both TM pilots and TS pilots are jointly employed for channel estimation.
\vspace{-3mm}
\subsection{Hybrid Pilot Framework}
\vspace{-1mm}
Without loss of generality, we consider a frame-based transmission, where each frame comprises a training overhead of $\tau$ pilots and $T-\tau$ data symbols. The framework of the proposed hybrid pilots is shown in Fig. 1, where the training overhead is composed of $(1-\alpha)\tau$ TM pilots, followed by $\alpha\tau$ TS pilots. $\alpha \in [0,1]$ denote the time fraction allocated between TM pilots and TS pilots. Note that $\alpha \to 0$ and $\alpha \to 1$ denotes that either TM pilots or TS pilots are deployed in training overhead. Therefore, the conventional methods employing only TM pilots or TS pilots is a special case of the proposed scheme.

\begin{figure} [!t]
\centering
\includegraphics[width=3.0in]{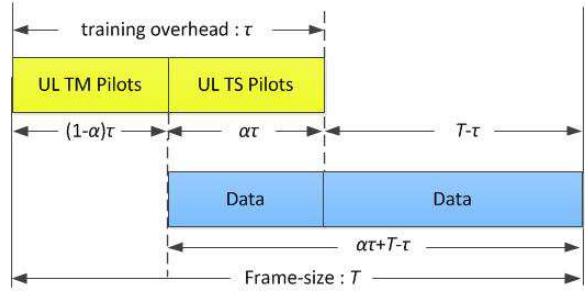}
\vspace{-3mm}
\caption{UL frame structure of a hybrid pilot-aided system.}
\vspace{-10mm}
\label{fig0}
\end{figure}

Denote $s_{j,k}(t)$, $p_{j,k}(t)$ are data and pilot symbols of the specific $k$-th user in cell $j, j = 1,2,\cdots,L$ at time $t$, respectively, the transmitted signal within the interval of training overhead has the form
\vspace{-5mm}
\begin{align} \label{mixX}
\textbf{x}_{j,k}= \textbf{s}_{j,k}+\textbf{p}_{j,k},
\vspace{-5mm}
\end{align}
where $\textbf{s}_{j,k}$ and $\textbf{p}_{j,k}$ are data and hybrid pilot vectors, respectively, given by
\vspace{-3mm}
\begin{align}
\textbf{s}_{j,k} &= [0,\cdots,0, s_{j,k}((1-\alpha) \tau+1),\cdots,s_{j,k}(\tau)]^T,
\end{align}
\begin{align}
\textbf{p}_{j,k} &= [\underbrace{p_{j,k}(1),\cdots,p_{j,k}((1-\alpha) \tau)}_{\textrm{TM Pilots}}, \underbrace{p_{j,k}((1-\alpha) \tau+1),\cdots,p_{j,k}(\tau)}_{\textrm{TS Pilots}}]^T.
\end{align}
Similar to \cite{TCOM_SP10} and \cite{TVT_SP09}, we assume that $\{s_{j,k}(t)\}, \forall j$ contain independent identically distributed (i.i.d.) samples, and are mutually independent to $\{p_{j,k}(t)\}$. The power of data
sequence and pilot sequence, respectively, are given by
\vspace{-2mm}
\begin{align} \label{HP}
    E\{|s_{j,k}(t)|^2\} &= 1-\lambda,~~~~~~  \\
    E\{|p_{j,k}(t)|^2\} &= \lambda,
    \vspace{-3mm}
\end{align}
where $\lambda \in (0,1)$ is the power-allocation factor between pilots and data.

\vspace{-3mm}
\subsection{Hybrid Pilot-aided Channel Estimation}
From (\ref{mixX}), the received signal matrix at the BS in cell 1, denoted by, $\textbf{Y}_1 \in \mathcal{C}^{M \times \tau}$, has the form
\begin{align}
\textbf{Y}_{1} = \sum_{k=1}^{K}\textbf{h}_{1,k}\big(\textbf{p}_{1,k}+\textbf{s}_{1,k}\big) + \sum_{j \neq 1}^{L}\sum_{k=1}^{K}\textbf{h}_{j,k}\big(\textbf{p}_{j,k} + \textbf{s}_{j,k}\big) + \textbf{N}_{1}.
\end{align}
Treating data as interference, the channel estimates on $\hat{\textbf{h}}_{1}$ of the $k$-th UT in the target cell can be obtained using LS criteria
\begin{equation}
\label{LS}
\hat{\textbf{h}}_{1,k} = \textrm{arg} \mathop{\min}_{\textbf{h}} \big\|\textbf{Y}_{1} - \textbf{h}_{1,k}\textbf{p}_{1,k} \big\|^2_F.
\end{equation}
Since the hybrid pilots comprise TS pilots superimposed onto the data symbols, the size of training overhead can be much longer than that employing only TM pilots. Therefore, provided that $\tau \geq KL$\footnote{For high mobility case of $T \leq KL$, one can resort to channel modeling (i.e., basis expansion model) to reduce the channel unknowns, and then employ the two-step procedure to obtain channel estimates over multiple UL frames\cite{TCOM_SP10}. The details for the analysis of $T \leq KL$ is omitted herein due to space constraint.}, each user can be assigned a unique orthogonal pilot to avoid pilot contamination while preserving transmission efficiency. Accordingly, the estimation on $\textbf{h}_{1,k}$ is given by
\begin{small}
\begin{align}
\label{Hhat}
\hat{\textbf{h}}_{1,k} = \textbf{h}_{1,k}+ \underbrace{\frac{1}{\|\textbf{p}_{1,k}\|^2}\sum_{k'=1}^{K}\textbf{h}_{1,k'}\big( \textbf{s}_{1,k'}\textbf{p}_{1,k}^{H} \big)+\frac{1}{\|\textbf{p}_{1,k}\|^2} \sum_{j = 1}^L \sum_{k'=1}^{K}\textbf{h}_{j,k'} \big(\textbf{s}_{j,k'}\textbf{p}_{1,k}^{H} \big) + \frac{1}{\|\textbf{p}_{1,k}\|^2} \textbf{N}_{1}\textbf{p}_{1,k}^{H} }_{\Delta \textbf{h}_{1,k}}.
\end{align}
\end{small}
\!\!In the above, $\Delta {\textbf{h}}_{1,k}$ is the interference to channel estimation. To measure the estimation quality in (\ref{Hhat}), we derive the normalized channel mean square error (MSE) on $\hat{\textbf{h}}_{1,k}$ as
\begin{align} \label{MSE}
\sigma_{\Delta \textbf{h}_{1,k}}^2 = \frac{E\big[\|\Delta \textbf{h}_{1,k}\|^2 \big]}{E\big[\|\textbf{h}_{1,k}\|^2\big]} = \frac{1}{\tau} \bigg( \frac{\alpha(1-\lambda)}{\lambda} \frac{\sum_{j=1}^L\sum_{k=1}^{K}\beta_{j,k}}{\beta_{1,k}} + \frac{\sigma_n^2}{\lambda \beta_{1,k}} \bigg).
\end{align}
As shown in (\ref{MSE}), although orthogonal pilots have been assigned to users to eliminate pilot contamination, the performance of channel estimation suffers from data interference, and thereby, is inversely proportional to the time ratio and power of data $\alpha (1-\lambda)$. Besides, $\sigma_{\Delta \textbf{h}_{1,k}}^2$ reduces linearly with $\tau$ and $\lambda$. This is expected, since $\tau$ independent pilots are involved for channel estimation. Later we will see that the hybrid pilot-based scheme with optimized $\alpha$, $\lambda$ and $\tau$ yields a substantial improvement in
the average achievable rate.

\vspace{-5mm}
\subsection{Analysis of Achievable UL Rate}
From (\ref{mixX}), when $\alpha \neq 0$, the interval of training overhead also contains part of data symbols $s_{j,k}(t), t=(1-\alpha)\tau + 1, \cdots, \tau$. Thus, we perform data detection at $k$-th UT in target cell in two phases, i.e., Phase 1): data phase mixed with TS pilots (of a size $\alpha \tau$), and Phase 2): pure data phase of $T-\tau$ symbols.

Recall (\ref{rxvector}), the received signal of the above two phases at time $t$, denoted by $\textbf{y}_1^{\textrm{I}}(t), t = (1-\alpha)\tau + 1,\cdots,\tau$ and $\textbf{y}_1^{\textrm{II}}(t), t=\tau+1,\cdots,T$, respectively, can be written in signal-plus-interference forms,
\begin{subequations}
\begin{align}
\label{detection}
\textbf{y}_1^{\textrm{I}}(t) &= \textbf{y}_1(t)-\hat{\textbf{h}}~\!\!^H_{1,k}  p_{1,k}(t) \nonumber \\
&= \hat{\textbf{h}}_{1,k} s_{1,k}(t) + \Delta\hat{\textbf{h}}_{1,k} x_{1,k}(t) + \sum_{k'\neq k}^{K}\textbf{h}_{1,k'}x_{1,k'}(t) + \sum_{j\neq 1}^L\sum_{k'=1}^{K}\textbf{h}_{j,k'}x_{j,k'}(t) + \textbf{n}_{1}(t) \\
\textbf{y}_1^{\textrm{II}}(t) &= \hat{\textbf{h}}_{1,k} s_{1,k}(t) + \Delta\hat{\textbf{h}}_{1,k} s_{1,k}(t) + \sum_{k'=1}^{K}\textbf{h}_{1,k'} s_{1,k'}(t) + \sum_{j\neq 1}^L\sum_{k'=1}^{K} \textbf{h}_{j,k'} s_{j,k'}(t) + \textbf{n}_{1}(t).
\end{align}
\end{subequations}
To maintain low receiver complexity, we employ a simple MF detector. The detected output are respectively given by
\begin{subequations}
\begin{align}
\label{detection1}
\hat{\textbf{h}}~\!\!^H_{1,k} \textbf{y}_1^{\textrm{I}}(t) =& \underbrace{\| \hat{\textbf{h}}_{1,k} \|^2 s_{1,k}(t)}_{S(t)} + \underbrace{\hat{\textbf{h}}~\!\!^H_{1,k} \Delta\hat{\textbf{h}}_{1,k} x_{1,k}(t)}_{I_{1}(t)} +\sum_{k'\neq k}^{K}\underbrace{\hat{\textbf{h}}~\!\!^H_{1,k}\textbf{h}_{1,k'} x_{1,k'}(t)}_{I_{2}(t)} \nonumber \\
&+\sum_{j\neq 1}^L \sum_{k'=1}^{K}\underbrace{\hat{\textbf{h}}~\!\!^H_{1,k} \textbf{h}_{j,k'} x_{j,k'}(t)}_{I_{3}(t)}+ \underbrace{\hat{\textbf{h}}~\!\!^H_{1,k} \textbf{n}_{1}(t)}_{N(t)},  \\
\label{detection2}
\hat{\textbf{h}}~\!\!^H_{1,k} \textbf{y}_1^{\textrm{II}}(t) =& \underbrace{\| \hat{\textbf{h}}_{1,k} \|^2 s_{1,k}(t)}_{S(t)} + \underbrace{\hat{\textbf{h}}~\!\!^H_{1} \Delta\hat{\textbf{h}}_{1,k} s_{1,k}(t)}_{I_{1}^{\prime}(t)} +\sum_{k'\neq k}^{K}\underbrace{\hat{\textbf{h}}~\!\!^H_{1,k}\textbf{h}_{1,k'} s_{1,k'}(t)}_{I_{2}^{\prime}(t)} \nonumber \\
&+\sum_{j\neq 1}^L\sum_{k'=k}^{K} \underbrace{\hat{\textbf{h}}~\!\!^H_{1,k} \textbf{h}_{j,k'} s_{j,k'}(t)}_{I_{3}^{\prime}(t)} + \underbrace{\hat{\textbf{h}}~\!\!^H_{1,k} \textbf{n}_{1}(t)}_{N(t)}.
\end{align}
\end{subequations}
In the given equations, the first terms on right-hand-side $S(t)$ are the desired signals, while the rest four terms are attributed to interference. In particular, since the detection is based on $\hat{\textbf{h}}~\!\!^H_{1,k}$, we treat $I_1(t)$ and $I_1^{\prime}(t)$ as interference, although both terms contain part of the desired signal. Therefore, we refer to $I_1(t)$ and $I_1^{\prime}(t)$ as self-interference. By similarity, we refer to $I_2(t)$, $I_2^{\prime}(t)$, $I_3(t)$ and $I_3^{\prime}(t)$ as cross-interference since these terms contain interference across all $L$ cells.

According to Jensen$'$s inequality, a lower bound on the achievable uplink rate of $k$-th UT can be written as
\begin{align}
\label{lowerboundR}
R_{1,k}\geq\widetilde{R}_{1,k}=\textrm{log}_2\bigg(1+\frac{1}{E\big[\frac{1}{\gamma}\big]}\bigg),
\end{align}
where $\gamma$ is the signal-to-interference-plus-noise ratio (SINR). From (\ref{detection1}) and (\ref{detection2}), the UL rate for using hybrid pilot-aided channel estimation is lower-bounded
\begin{align}
\label{throughput}
\widetilde{R}_{1,k}(\alpha, \tau, \lambda) = \frac{\alpha\tau}{T} \textrm{log}_2\bigg(1 + \frac{1}{E\big[\frac{1}{\gamma^{\textrm{I}}}\big]}\bigg) + \Big(1-\frac{\tau}{T}\Big) \textrm{log}_2\bigg(1+\frac{1}{E\big[\frac{1}{\gamma^{\textrm{II}}}\big]}\bigg),
\end{align}
where $\gamma^{\textrm{I}}$ and $\gamma^{\textrm{II}}$ are respectively the SINRs contained in the output of the MF detector in (\ref{detection1}) and (\ref{detection2}), which can be expressed as
\begin{subequations}
\begin{align}
\label{SINR}
\gamma^{\textrm{I}} &= \frac{E\big[|S(t)|^2\big]}{E\big[|I_1(t)|^2\big]+\sum_{k'\neq k}^{K}E\big[|I_2(t)|^2\big] +\sum_{j\neq 1}^{L}\sum_{k'\neq k}^{K}E\big[|I_3(t)|^2\big]+E\big[|N(t)|^2\big]}, \\
\label{SINR1}
\gamma^{\textrm{II}} &= \frac{E\big[|S(t)|^2\big]}{E\big[|I_1^{\prime}(t)|^2\big]+\sum_{k'\neq k}^{K}E\big[|I_2^{\prime}(t)|^2\big] + \sum_{j\neq 1}^{L}\sum_{k'\neq k}^{K}E\big[|I_3^{\prime}(t)|^2\big]+E\big[|N(t)|^2\big]}.
\end{align}
\end{subequations}
\newtheorem{lemma}{Lemma}
\begin{lemma}
For fixed values of $\alpha$ and $\lambda$, when $M$ is large, the approximate SINR in (\ref{SINR}) and (\ref{SINR1}), denoted by $\gamma^{\textrm{I}}_{app}$ and $\gamma^{\textrm{II}}_{app}$, respectively, are given by
\end{lemma}
\begin{subequations}
\vspace{-10mm}
\begin{align}
\label{SINR_app}
\gamma^{\textrm{I}}_{app} &= \frac{(1-\lambda) \beta_{1,k}^2}{\frac{(1-\lambda)\alpha}{\lambda} \frac{1}{\tau}b_1 + \frac{1}{M} \Big(b_2 + \sigma_n^2\beta_{1,k}\Big)}, \\
\label{SINR1_app}
\gamma^{\textrm{II}}_{app} &= \frac{(1-\lambda)\beta_{1,k}^2} {\frac{(1-\lambda)^2\alpha}{\lambda} \frac{1}{\tau} b_1 + \frac{1}{M} \Big((1-\lambda)b_2 + \sigma_n^2\beta_{1,k} \Big)},
\end{align}
\end{subequations}
where $b_1=\sum_{j = 1}^L\sum_{k=1}^{K} \beta_{j,k}^2$ and $b_2=\sum_{j\neq 1}^{L}\sum_{k=1}^{K} \beta_{1,k} \beta_{j,k}+\sum_{k'\neq k}^{K}\beta_{1,k}\beta_{1,k'}$.
\begin{proof}
See Appendix-A.
\end{proof}

Substituting (\ref{SINR_app}) and (\ref{SINR1_app}) into (\ref{throughput}), the UL rate from UT k is given by
\begin{align}
\label{throughputasymp}
\widetilde{R}_{1,k}(\alpha, \tau, \lambda) \to &\frac{\alpha\tau}{T} \textrm{log}_2 \bigg(1+\frac{(1-\lambda) \beta_{1,k}^2}{\frac{(1-\lambda)\alpha}{\lambda} \frac{1}{\tau}b_1 + \frac{1}{M} \big(b_2 + \sigma_n^2\beta_{1,k}\big)} \bigg) \nonumber \\
&+ \Big(1-\frac{\tau}{T}\Big) \textrm{log}_2 \bigg(1 +\frac{(1-\lambda)\beta_{1,k}^2} {\frac{(1-\lambda)^2\alpha}{\lambda} \frac{1}{\tau}b_1 + \frac{1}{M} \big((1-\lambda)b_2 + \sigma_n^2\beta_{1,k} \big)} \bigg).
\end{align}
From (\ref{throughputasymp}), we have the following observations:
\begin{itemize}
  \item The UL rate from $k$-th UT for employing hybrid pilots is limited even when $M \to \infty$, and can be well approximated for large $M$ as
        \begin{align}
            {\lim_{M\to \infty}} \widetilde{R}_{1,k}(\alpha, \tau, \lambda) \to& \frac{\alpha\tau}{T} \textrm{log}_2 \bigg(1+\frac{\beta_{1,k}^2}{\frac{\alpha}{\lambda \tau}b_1 } \bigg)+ \Big(1-\frac{\tau}{T}\Big) \textrm{log}_2 \bigg(1+\frac{ \beta_{1,k}^2} {\frac{\alpha(1-\lambda)}{\lambda \tau}b_1 } \bigg).
        \end{align}
        The result implies that, although the hybrid pilot-based scheme cannot completely mitigate the effect of pilot contamination, it provides the potential for significant improvement in the ultimate performance in comparison with the conventional pilot-aided designs through the following three adjustment factors: (1) The time-ratio between TM pilots and TS pilots $\alpha$, (2) the time allocated to hybrid pilots (training overhead) ${\tau}$, and (3) the power-ratio between pilots and data $\lambda$.
        It will be shown in Section VI numerical results that the hybrid pilot-aided scheme with optimal $\alpha$, $\lambda$ and $\tau$ yields a substantial improvement in the UL achievable rate. It is also worth noting that the conventional TS pilot \cite{SP1,TVT16} is in principle the special case of the hybrid pilot when $\tau \to T$ and $\alpha \to 1$.
\end{itemize}

\vspace{-3mm}
\subsection{Performance Enhancement: A Data-aided Solution}
Denote $\hat{s}_{j,k}(t)$ and $\Delta {s}_{j,k}(t)$ as the detection and the detection error after hard-decision operation w.r.t. the data symbol $s_{j,k}(t)$ of the target user in cell $j, j = 1,2,\cdots,L$ obtained by using (\ref{LS}) and MF detector (\ref{detection1}). Then, we have
\begin{align}
\Delta {s}_{j,k}(t) = s_{j,k}(t) - \hat{s}_{j,k}(t), ~k=1,2,\cdots,K.
\end{align}
As pointed out in \cite{DA14}, we make the following assumptions:
\begin{enumerate}
  \item Both $\{\hat{s}_{j,k}(t)\}$ and $\{\Delta {s}_{j,k}(t)\}$ are zero-mean and contain i.i.d. samples,
  \item $\{\Delta s_{j,k}(t)\}$ and $\{s_{j,k}(t)\}$ are mutually independent\footnote{We show in Appendix A-D that the correlation between self-interference $I_1(t)$ and the desired signal $S(t)$ in (15a) is inversely proportional to the training size $\tau$. Thereby, assumption 2) is fairly accurate in scenarios with large values of $\tau$ and $M$.}.
\end{enumerate}
For a given signal constellation, e.g., $M$-PSK ($M = 2, 4, 8, \cdots$), and consider the worst case by assuming the farthest neighbor selection when executing data decoding. Define the distance between the data signal in the target cell and its detected data as $d_k(t) = |s_{1,k}(t) - \hat{s}_{1,k}(t)|^2$. Suppose hard decision is employed, then we have
\begin{align}
d_k(t) = |\Delta s_{1,k}(t)| =
\begin{cases}
~~~~ 0, ~~ &\textrm{w.p.}~ 1-p_{e,k} \\
2\sqrt{1-\lambda}, ~~&\textrm{w.p.}~~~p_{e,k}
\end{cases}
\end{align}
leading to
\begin{align}
E\big[\Delta s_{1,k}(t_1)\Delta s_{1,k}^{*}(t_2)\big] = 4(1-\lambda)\delta(t_1-t_2),
\end{align}
where $p_{e,k}$ is the steady-state error probability of data decoding at the BS w.r.t. the user in the target cell.

The detected data symbols and the estimated channel of the desired user are then used in feedback to iteratively refine the estimation quality by mitigating correlation between TS pilots and data in (\ref{LS}). To be specific, denote $\hat{\textbf{h}}_{1,k}^{(i)}$ as the corresponding channel estimates of the $i$-th iteration for using the iterative data-aided solution, we have
\begin{small}
\begin{align}
\hat{\textbf{h}}_{1,k}^{(i)} &=\Big(\textbf{Y}_{1}-\sum_{k'=1}^{K}\hat{\textbf{h}}~\!\!^{(i-1)}_{1,k'}\hat{\textbf{s}}_{1,k}\Big)\cdot\frac{\textbf{p}_{1,k}^H }{\|\textbf{p}_{1,k}\|^2} \nonumber \\
& = \textbf{h}_{1,k} + \underbrace{\frac{1}{\|\textbf{p}_{1,k}\|^2} \bigg(\sum_{k'=1}^{K}\Big(\textbf{h}_{1,k'} \big( \Delta \textbf{s}_{1,k'}\textbf{p}_{1,k}^{H} \big) + \Delta \textbf{h}_{1,k}^{(i-1)} \big(\hat{\textbf{s}}_{1,k}\textbf{p}_{1,k}^{H} \big)\Big) + \sum_{j\neq 1}^L\sum_{k'=1}^{K}\textbf{h}_{j,k} \big( \textbf{s}_{j,k'}\textbf{p}_{1,k}^{H} \big) + \textbf{N}_{1}\textbf{p}_{1,k}^{H} \bigg)}_{\Delta \textbf h^{(i)}_{1,k}}.
\end{align}
\end{small}
$\!\!$where $\Delta\textbf{h}_{1,k}^{(i-1)} = \textbf{h}_{1,k} - \hat{\textbf{h}}~\!\!_{1,k}^{(i-1)}$. Using the property $\frac{1}{\tau}E[|\Delta \textbf{s}_{1,k}\textbf{p}_{1,k}^{H} |^2] \to 4 p_{e,k} \lambda(1-\lambda) $, and assuming that $\Delta\textbf{h}_{1,k}^{(i)} \approx \Delta\textbf{h}_{1,k}^{(i-1)}$ when $i$ is large, we perform data detection of the $i$-th iteration as
\begin{align}
\hat{s}_{1,k}^{(i)}(t) \!=\!
\begin{cases}
s_{1,k}(t) \!+\! \frac{(\hat{\textbf{h}}~\!\!^{(i)}_{1,k})^H}{\| \hat{\textbf{h}}~\!\!^{(i)}_{1,k} \|^2 } \big(\sum_{k'=1}^{K}\Delta\hat{\textbf{h}}_{1,k'} x_{1,k'}(t) \!+\! \sum_{j\neq 1}^L \sum_{k'=1}^{K}\textbf{h}_{j,k'} x_{j,k'}(t) + \textbf{n}_{1}(t) \big),\\ ~~~~~~~~~~~~~~~~~~~~~~~~~~~~~~~~~~~~~~~~~~~~~~ t = (1-\alpha)\tau + 1,\cdots,\tau,
\\
s_{1,k}(t) \!+\! \frac{(\hat{\textbf{h}}~\!\!^{(i)}_{1,k})^H}{\| \hat{\textbf{h}}~\!\!^{(i)}_{1,k} \|^2 } \big(\sum_{k'=1}^{K}\Delta\hat{\textbf{h}}_{1,k'} s_{1,k'}(t) \!+\! \sum_{j\neq 1}^L \sum_{k'=1}^{K}\textbf{h}_{j,k'} s_{j,k'}(t) + \textbf{n}_{1}(t) \big), \\ ~~~~~~~~~~~~~~~~~~~~~~~~~~~~~~~~~~~~~~~~~~~~~~~~~t = \tau + 1,\cdots,T.
\end{cases}
\end{align}
$\!\!$Simplifying the resulting expressing similarly as in the case of initial MF detection in Section IV-B, the achievable UL rate of $i$-th iteration at $k$-th UT in the target cell is given by
\begin{align}
\label{throughputasymp_ith}
\widetilde{R}_{1,k}^{(i)}(\alpha, \tau, \lambda) =&
\frac{\alpha\tau}{T} \textrm{log}_2 \bigg(1+\frac{(1-\lambda) \beta_{1,k}^2}{\frac{(1-\lambda)\alpha}{\lambda \tau}c_1+ \frac{1}{M} \big(c_2+ \beta_{1,k} \sigma_n^2\big) } \bigg) \nonumber \\
&+ \Big(1-\frac{\tau}{T}\Big) \textrm{log}_2 \bigg(1 + \frac{(1-\lambda)\beta_{1,k}^2} {\frac{(1-\lambda)^2\alpha}{\lambda \tau}c_1 + \frac{1}{M} \big((1-\lambda)c_2+ \beta_{1,k} \sigma_n^2\big)} \bigg).
\end{align}
where $c_1=\sum\limits_{k'=1}^{K}4p_{e,k} \cdot \beta_{1,k'}^2 +\sum\limits_{j \neq 1}^L\sum\limits_{k'=1}^{K} \beta_{j,k'}^2$ and $c_2=\sum\limits_{j\neq 1}^{L}\sum\limits_{k'=1}^{K} \beta_{1,k }\beta_{j,k'} +\sum\limits_{k'\neq k}^{K}4p_{e,k'}\beta_{1,k}\beta_{1,k'}$.

In summary, the iterative data-aided solution can be interpreted as follows: In each step of iteration, hybrid pilots are used to estimate the channel, by which data detection is obtained through a MF detection. Then, the detected data is employed to refine the channel estimates by mitigating the correlation between data and pilots of the desired user within the TS pilots phase, and in turn, to improve the quality of channel and data estimation in the forthcoming step, and thereby the UL achievable rate.

\section{Asymptotically Optimal Solutions}
In this section, we optimize the variables, including the time-ratio between TM and TS pilots $\alpha$, the power-ratio between pilots and data $\lambda$, and the size of training overhead $\tau$, aiming to improve the ultimate UL rate.

\vspace{-3mm}
\subsection{Problem Formulation}
Achieving user fairness, in order to maximize the minimum rate for all users $\widetilde{R}_{1,k}^{(i)}(\alpha, \tau, \lambda)$ in (\ref{throughputasymp_ith}), we have the following problem formulation
\begin{align}
\label{optimization}
(\textmd{P}1):  \max_{\alpha,\lambda,\tau} ~~~~ & \min_{1\leq k\leq K}\widetilde{R}_{1,k}^{(i)}(\alpha,\lambda,\tau),  \\
\textrm{s. t.}~~~~ & 0 \leq \alpha \leq 1, \nonumber \\
& 0  \leq \lambda \leq 1,  \nonumber \\
& KL \leq \tau \leq T.  \nonumber
\end{align}
For fixed $M$, the direct optimization on $(\textmd{P}_1)$ is challenging due to the nonlinear relationship between the UL rate and the variables $\alpha$, $\lambda$ and $\tau$. In massive MIMO systems, these variables are coupled between the training phase and the data phase, i.e.,
\begin{itemize}
  \item In training phase, estimation quality depends on the $\alpha$, $\tau$, as well as the power-ratio $\lambda$.
  \item The estimation quality affects the detection performance and the UL rate.
  \item Besides the estimation quality, the UL rate depends on the ratio of data phase over the frame, which depends on both $\alpha$ and $\tau$.
\end{itemize}
Nevertheless, we can obtain interesting asymptotical solutions and insights in the large-$M$ regime.

\vspace{-3mm}
\subsection{Asymptotical Optimization on Time-Ratio between TM and TS Pilots}
\begin{lemma}
The convexity or concavity of $\widetilde{R}_{1,k}$ in (P1)\footnote{For simplicity, we drop the superscript of $\widetilde{R}_{1,k}^{(i)}(\alpha,\lambda,\tau)$ in (\ref{throughputasymp_ith}), and denote it by $\widetilde{R}_{1,k}$ in the following of this paper.} w.r.t. $\alpha$ depends on the time-ratio of training overhead over the whole frame $\frac{\tau}{T}$, which includes the following three cases:
\begin{itemize}
  \item Case 1: When $\frac{\tau}{T} \in (0, \frac{1}{2+2g})$, $\widetilde{R}_{1,k}$ is a convex function w.r.t. $\alpha$.
  \item Case 2: When $\frac{\tau}{T} \in (\frac{1}{1+2g},1)$, $\widetilde{R}_{1,k}$ is a concave function w.r.t. $\alpha$.
  \item Case 3: When $\frac{\tau}{T} \in [\frac{1}{2+2g}, \frac{1}{1+2g}]$, $\widetilde{R}_{1,k}$ is a convex function when $\alpha \in [0,\frac{T}{\tau}-1-2g]$, and a concave function when $\alpha \in [\frac{T}{\tau}-1-2g,1]$, respectively, where $g= \frac{\lambda\tau(c_2 + \beta_{1,k} \sigma_n^2)}{M(1-\lambda)c_1}$.
\end{itemize}
\end{lemma}
\begin{IEEEproof}
See Appendix B-A.
\end{IEEEproof}
From lemma 2, the derivative of $\widetilde{R}_{1,k}$ w.r.t. $\alpha$ depends on $\frac{\tau}{T}$ and $g$. Thus, the direct optimization on $\alpha$ is challenging since exhaustive search is of high complexity. To this end, we firstly evaluate the monotonicity of $\widetilde{R}_{1,k}$ w.r.t. $\alpha$, and then propose an iterative bisection procedure as follows.

{
\small{
\vspace*{-3mm}
\noindent\hrulefill{\hrule height 0.6pt}
\vspace*{-2mm}
\begin{center}
   \textbf{Algorithm 1}: Optimization of Time-Ratio between TM Pilots and TS Pilots
\end{center}
\vspace*{-8mm}
\noindent\hrulefill{\hrule height 0.8pt}
\vspace*{1mm}
\begin{algorithmic}[1]
\REQUIRE The parameter from (\ref{throughputasymp_ith})
\ENSURE Optimal Time-Ratio between TM Pilots and TS Pilots $\alpha$
\STATE Calculate $\frac{1}{1+2g}$, $\frac{1}{2+2g}$ and $\frac{T}{\tau}-1-2g$ as mentioned in Lemma 2. If $\frac{\tau}{T}<\frac{1}{2+2g}$, go to 2. Otherwise, go to 3.
\STATE Set $\alpha^{opt}$ = arg~max\{$\widetilde{R}_{1,k}(0),\widetilde{R}_{1,k}(1)$\}, since $\widetilde{R}_{1,k}$ is a convex function w.r.t. $\alpha$ (case 1);
\STATE Define the concave interval of $\widetilde{R}_{1,k}$ according to the value of $\frac{\tau}{T}$. If $\frac{\tau}{T}<\frac{1}{1+2g}$, we let $a=\frac{T}{\tau}-1-2g$ and $b=1$ (case 3), or Let $a=0$ and $b=1$ (case 2). Calculate $\widetilde{R}_{1,k}'(a)$ and $\widetilde{R}_{1,k}'(b)$ by (\ref{derTP}), since $\widetilde{R}_{1,k}'$ is a monotonically decreasing function w.r.t. $\alpha$. If $\widetilde{R}_{1,k}'(a)<0$, set $\alpha^{opt}=0$. If $\widetilde{R}_{1,k}'(b)>0$, set $\alpha^{opt}=1$. Otherwise, go to 4.
\STATE Start the iterative bisection procedure (For case 2 and 3 concave interval). Repeat the follow until $|\widetilde{R}_{1,k}'(\frac{a+b}{2})|<\epsilon$: Calculate $\widetilde{R}_{1,k}'(\frac{(a+b)}{2})$ and update the concave interval. If $\widetilde{R}_{1,k}'(\frac{a+b}{2})>0$, let $a=\frac{a+b}{2}$. Otherwise, let $b=\frac{a+b}{2}$. Set $\alpha^{opt}=\frac{a+b}{2}$. If $\frac{\tau}{T}<\frac{1}{1+2g}$ (case 3 concave interval), go to 5.
\STATE If $\widetilde{R}_{1,k}(\alpha^{opt})<\widetilde{R}_{1,k}(0)$, set $\alpha^{opt}=0$.
\end{algorithmic}
\vspace*{-5mm}
\noindent\hrulefill{\hrule height 0.6pt}
}
}

\subsection{Asymptotical Optimization on Power-Ratio Allocated to Pilots}
From (\ref{throughputasymp_ith}), the asymptotically optimal power-ratio $\lambda_{opt}$ is given in Lemma 3.
\vspace*{-2mm}
\begin{lemma}
For fixed $\alpha$ and $\tau$, the asymptotically optimal power-ratio between data and pilots is given by
\begin{align}
\label{lambdaopt}
\lambda^{opt}=\frac{(T-\tau)+2\alpha\tau-\sqrt{(T-\tau)^2+4\tau^2f\big((T-\tau)+\alpha\tau\big)}}{2(\alpha\tau-\tau^2f)},
\end{align}
\end{lemma}
where $f=\frac{c_2+ \beta_{1,k} \sigma_n^2}{Mc_1}$.
\begin{IEEEproof}
See Appendix B-B.
\end{IEEEproof}
From (\ref{lambdaopt}), it can be seen that, for arbitrary $\alpha$ and $\tau$, $\lambda^{opt}\rightarrow1$ when $M\rightarrow \infty$, as pointed out in \cite{TVT16}. A brief explanation for this behavior is that, $M \to \infty$ results in a nearly ``noise-free'' transmission background. It this case, increasing $\lambda$ can $always$ improve the estimation quality, and in turn, improves the UL rate without deteriorating the effective SNR.

\vspace{-3mm}
\subsection{Asymptotical Optimization on Time Allocated to Training Overhead}
The optimization on $\tau^{opt}$ is given in Lemma 4.
\begin{lemma}
The optimization on $\tau, \tau \in [KL,T]$, depends on both the power ratio of pilots $\lambda$ and the time ratio between TM pilots and TS pilots $\alpha$, which includes the following two cases,
\begin{itemize}
  \item Case 1: For an arbitrary $\alpha \in [0,1]$, when $\lambda \in (1- 2^{-\frac{T}{(1+KL h) KL ln2}},1]$, $\tau^{opt} = KL$.
  \item Case 2: For an arbitrary $\alpha \in [0,1]$, when $\lambda \in [0, 1- 2^{-\frac{T}{(1+KL h) KL ln2}}]$, $\tau^{opt} \in (KL, T]$, where in particular $\tau^{opt} = T$ when $\alpha = 1$ and $h = \frac{\lambda}{(1-\lambda)\alpha M}\frac{c_2 + \beta_{1,k} \sigma_n^2}{c_1}$.
\end{itemize}
\vspace*{-4mm}
\end{lemma}
\begin{IEEEproof}
See Appendix B-C.
\end{IEEEproof}
An important consequence of $Case~1$ in Lemma 4 is that, in scenarios of large pilot power where $\lambda$ is large enough to acquire a precise quality of channel estimation, reducing the training-size $\tau$ is beneficial to increase the efficiency of data transmission, as pointed out in \cite{PD3}. In contrast, for cases of negligible pilot power $\lambda$ (i.e., $Case~2$ of Lemma 4), where the estimation quality is not able to satisfy the detection quality, it is essential to increase $\tau$ to enhance the estimation performance, and in turn, to improve the cell rate. Particularly, we show in Appendix B-C that when $\{\lambda \in [0, 1-2^{-\frac{1}{\cdot(1+Th)ln2}}]\} \bigcup \{\alpha = [0,1)\}$, $\widetilde{R}_{1,k}$ is a concave function w.r.t. $\tau$, which implies that the global optimal $\tau^{opt}$ exists in the range $(KL, T]$. However, solving the optimization on $\tau$ is computational too expansive, because of the nonlinear relationship between $\alpha$ and $\tau$. To ease the computational burden, we firstly determine the monotonicity of $\widetilde{R}_{1,k}$ w.r.t. $\tau$, and then propose using an iterative bisection procedure to solve the optimization of $\tau$, see Algorithm 2.

{
\small{
\vspace*{-3mm}
\noindent\hrulefill{\hrule height 0.6pt}
\vspace*{-2mm}
\begin{center}
   \textbf{Algorithm 2}: Optimization of Time Allocated to Training Overhead
\end{center}
\vspace*{-8mm}
\noindent\hrulefill{\hrule height 0.6pt}
\vspace*{1mm}
\begin{algorithmic}[1]
\REQUIRE The parameter from (\ref{throughputasymp_ith})
\ENSURE Optimal time allocated to training overhead $\tau$
\STATE Calculate $\widetilde{R}_{1,k}'(KL)$ and $\widetilde{R}_{1,k}'(T)$ by (\ref{throughputderttau}), since $\widetilde{R}_{1,k}'$ is a monotonically decreasing function w.r.t. $\tau$. If $\widetilde{R}'_{1,k}(KL)<0$, set $\tau^{opt}=KL$. If $\widetilde{R}'_{1,k}(T)>0$, set $\tau^{opt}=T$. Otherwise, go to 2.
\STATE Let $a=KL$ and $b=T$, and start the iterative bisection procedure. Repeat the follow until $|\widetilde{R}_{1,k}'(\frac{a+b}{2})|<\epsilon$: Calculate $\widetilde{R}_{1,k}'(\frac{(a+b)}{2})$ and, update $a$ and $b$. If $\widetilde{R}_{1,k}'(\frac{a+b}{2})>0$, let $a=\frac{a+b}{2}$, or let $b=\frac{a+b}{2}$. Set $\tau^{opt}=\frac{a+b}{2}$
\end{algorithmic}
\vspace*{-5mm}
\noindent\hrulefill{\hrule height 0.6pt}
}
}


\section{Simulations}
Consider a cellular network with $L=7$ hexagonal cells, which consists of $1$ target cell and $6$ adjacent cells, with each cell $K=10$ users. The radius of each cell (from center to vertex) is normalized, and the users are assumed to be uniformly randomly distributed. We model the pass loss of a link from $k$-th user in cell $j$ to cell $i$ as $\beta_{j,k} = \big(\frac{d_{1,k}}{d_{j,k}}\big)^{\gamma} \beta_{1,k}$, where $d_{j,k}$ denotes the distance between the user of the target cell (cell $1$) and the BS of the $j$th cell, and $\gamma$ is the pass-loss exponent. We set $\gamma = 3.8$ and $\beta_{1,k} = 1$ for all k for simplicity as in \cite{HET13}, and assume the channel to be quasi-static during a frame of transmission $T$.

\vspace{-3mm}
\subsection{UL Achievable Rate}
Firstly, we conduct an experiment to validate the effectiveness of our theoretical analysis on the UL achievable rate described in Section III-C, where the average UL rate against different parameters (including the time-ratio $\alpha$, the power-allocation factors $\lambda$ and the size of training overhead $\tau$) are plotted in Fig. 2. The solid lines are obtained by approximations derived in (\ref{throughputasymp}) using Monte Carlo simulations. For reference, we also simulate the approximations derived in (\ref{throughputasymp}). As shown in Fig. 2, the agreement between the actual values and the approximated ones demonstrates the validity of our analysis. In addition, we note that the system performance gradually saturates as $M$ grows to infinity, i.e. $M \to 10^4$ in simulations, and the ultimate rate depends on the variables $\alpha, \lambda$ and $\tau$, as pointed out in the analysis in Section III-C.

\begin{figure} [!t]
\centering
\includegraphics[width=3.0in]{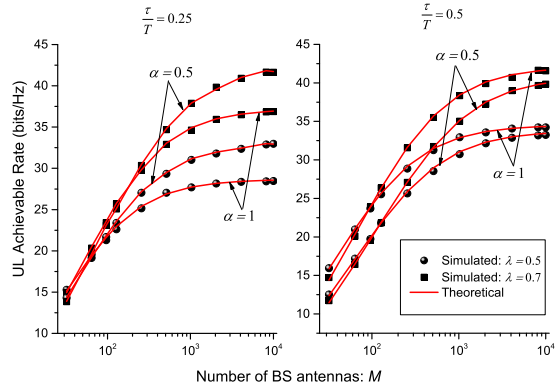}
\vspace{-5mm}
\caption{Theoretical v.s. simulated: UL rate for different variables $\alpha$, $\lambda$ and $\tau$ when SNR = 20dB.}
\vspace{-5mm}
\label{fig1}
\end{figure}

\begin{figure} [!t]
\centering
\includegraphics[width=3.0in]{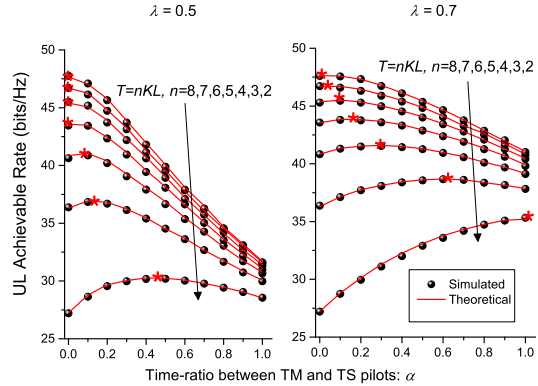}
\vspace{-5mm}
\caption{UL achievable rate for different values of $\alpha$ and $\tau$ for $\tau = KL$, and $\lambda = 0.5$ and $0.7$, respectively.}
\vspace{-4mm}
\label{fig2}
\end{figure}

\vspace{-3mm}
\subsection{Optimal Time-Ratio between TM Pilots and TS Pilots}
\begin{figure} [!t]
\centering
\includegraphics[width=3.0in]{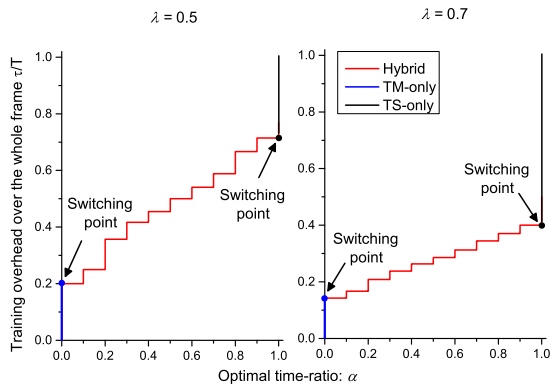}
\vspace{-5mm}
\caption{Optimal values of $\alpha$ for different values of $\frac{\tau}{T}$ ($M = 256$, $\lambda = 0.5, 0.7$ and SNR = 20dB).}
\vspace{-5mm}
\label{fig3}
\end{figure}

To explore the impact of time-ratio between TM pilots and TS pilots on the UL achievable rate, Fig. \ref{fig2} plots the UL achievable rates against the time-ratio $\alpha$, in order to validate the results described in Lemma 2. From Fig. \ref{fig2}, we can see that the optimal $\alpha$ depends on the time ratio $\frac{\tau}{T}$. Specifically, as the frame-size $T$ increases (which equivalent to $\frac{\tau}{T}$ reduces for fixed $\tau$), the optimal value of $\alpha$ reduces from 1 to 0. In particular, when $T$ is small, i.e. $\frac{\tau}{T} \to 1$, $\alpha^{opt} \to 1$, which indicates that the optimal design for hybrid training overhead contains only TS pilots. This fact implies that TS pilot is more suitable for transmission of a limited frame-size $T$. On the contrary, when $T$ is large, i.e., $\frac{\tau}{T} \to 0$, $\alpha^{opt} \to 0$. In this case, the optimal training overhead of the hybrid structure comprises only TM pilots, which implies that TM pilot is superior to TS pilot for large-frame based transmission. The above observations are also confirmed by the results shown in Fig. \ref{fig3}. Moreover, the tightness between the simulated results and the analytical ones further confirm the validity of our analysis.

\begin{figure} [!t]
\centering
\includegraphics[width=3.0in]{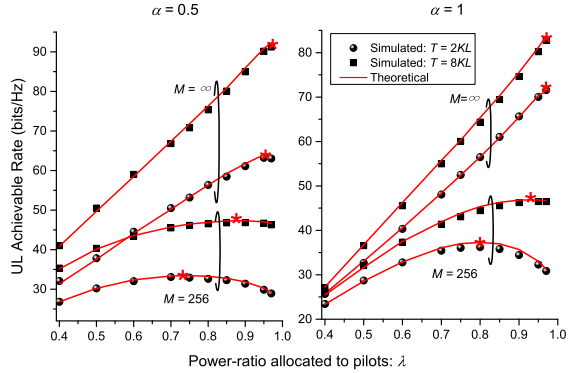}
\vspace{-5mm}
\caption{UL achievable rate versus different power-ratio between pilots and data $\lambda$ when SNR = 20 dB.}
\vspace{-7mm}
\label{fig4}
\end{figure}

\subsection{Optimal Power-Ratio Allocated to Pilots}

Next, we examine the effect of the power-ratio $\lambda$ on the UL rate, in order to validate our analysis in Lemma 3. It can be seen in Fig. \ref{fig4} that for finite $M$ (e.g. $M = 64, 256$) and fixed values of $\alpha$ (e.g. $\alpha = 0.5$ and $1$, respectively), the optimal values of $\lambda$ that maximize $R_1$ in (\ref{optimization}) are approximately $0.81$ and $0.85$ when $M = 256$ and $\alpha = 0.5$ and $1$, respectively. When $M \to \infty$ (we set $M = 10^4$ in simulations), $\lambda^{opt} \to 1$. The result is expected and can be explained as follows,
\begin{enumerate}
  \item For finite $M$, a larger $\lambda$ leads to better estimation quality but simultaneously reduces SNR. This fact deteriorates the system performance.
  \item When $M \to \infty$, the thermal noise vanishes due to the significant array gain. In such a ``noise-free'' scenario, increasing the $\lambda$ always improves the estimation quality without reducing SNR. This leads to an increased performance directly proportional to $\lambda$.
\end{enumerate}
The above results are consistent with our theoretical analysis detailed in Lemma 3. In addition, many of the simulation results generated in the course of this study (which have been removed here due to space constraints) also confirm that an excellent agreement exists between the actual values and the approximated ones.

\subsection{Optimal Time Allocation to Training Overhead}

\begin{figure} [!t]
\centering
\includegraphics[width=3.0in]{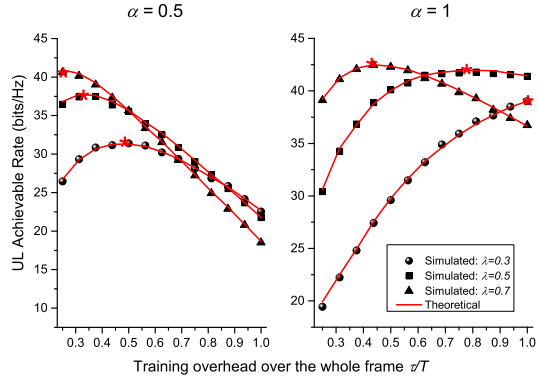}
\vspace{-5mm}
\caption{UL achievable rate versus different time-ratio of training overhead $\frac{\tau}{T}$ for different values of $\lambda$ when $\alpha = 0.5$ and $1$, respectively ($M$ = 256 and SNR = 20 dB).}
\vspace{-2mm}
\label{fig5}
\end{figure}

\begin{figure} [!t]
\centering
\includegraphics[width=3.0in]{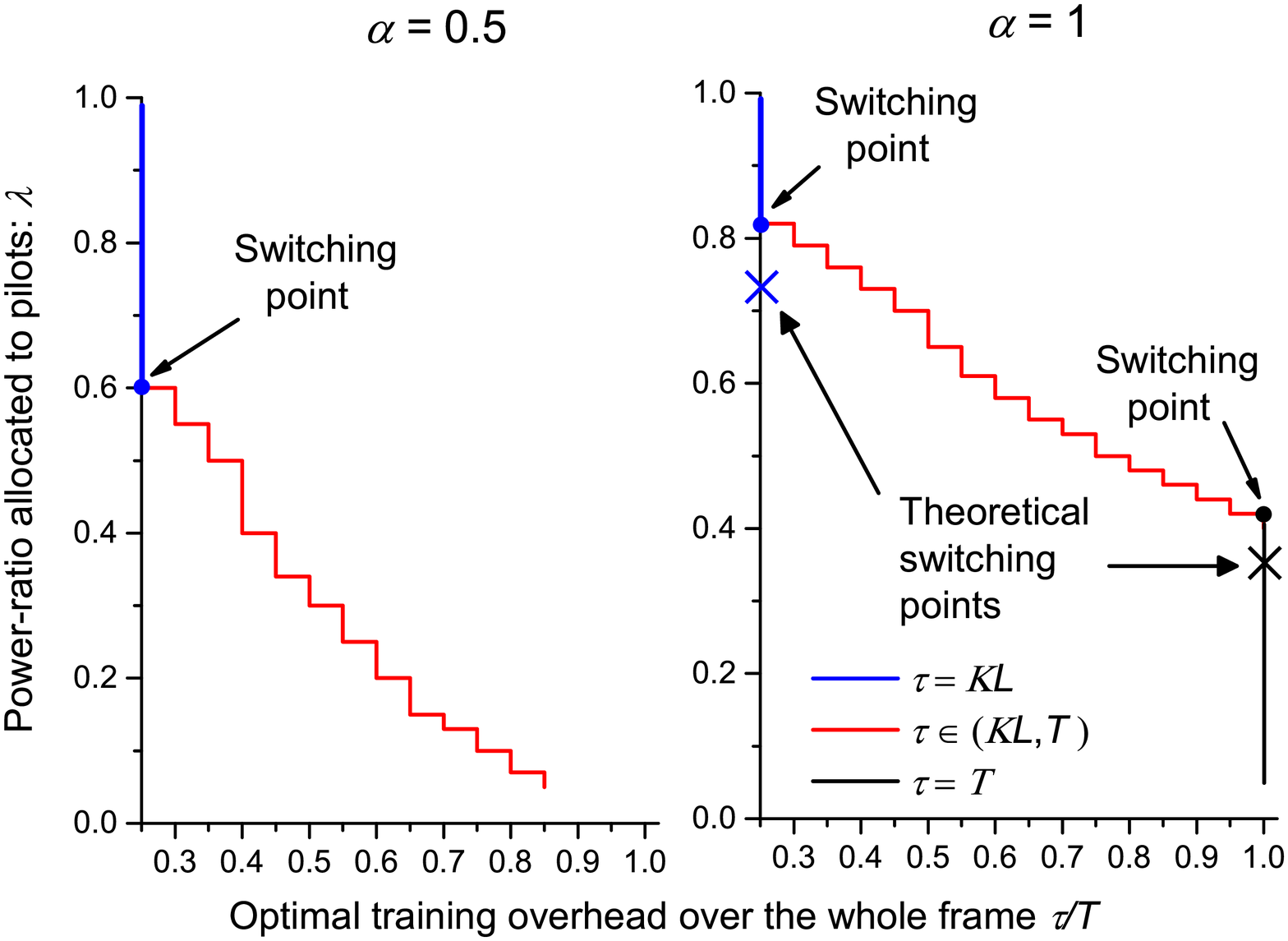}
\vspace{-5mm}
\caption{Optimal time-ratio of training overhead over the whole frame $\frac{\tau}{T}$ ($M$ = 256 and SNR = 20 dB).}
\vspace{-2mm}
\label{fig6}
\end{figure}

We now move forward to investigate the performance of UL rate versus the time-ratio of training overhead $\frac{\tau}{T}$. In Fig. \ref{fig5}, we observe that, for fixed $\alpha$, the optimal value of $\frac{\tau}{T}$ depends on the power ratio allocated to pilots $\lambda$. We also plot the optimal $\frac{\tau}{T}$ for different $\lambda$ in Fig. \ref{fig6}. The numerical results in both Fig. \ref{fig5} and Fig. \ref{fig6} agree with our analysis in Lemma 4 that $\tau \to KL$ when $\lambda$ is large, whereas ${\tau} \to T$ for small value of $\lambda$ when $\alpha = 1$. The explanation for this behavior is that, according to (\ref{MSE}), when $\lambda$ is small and not enough to acquire accurate channel estimates, it is essential to increase the training-size $\tau$ (i.e. $\frac{\tau}{T} \to 1$) to improve the estimation quality, in order to ensure the detection performance. In contrast, with the hybrid property in (\ref{HP}), increasing $\tau$ improves the estimation performance at the penalty of either introducing data interference or reducing the effective data rate. Thus, when $\lambda$ is large enough to satisfy (\ref{lambda_cond}), the gain of channel estimation by increasing $\tau$ is insufficient to compensate for the corresponding loss of rate. This fact leads to $\tau^{opt} \to KL$.

\subsection{Performance Comparison with the Conventional Pilot-based Schemes}

\begin{figure} [!t]
\centering
\includegraphics[width=3.0in]{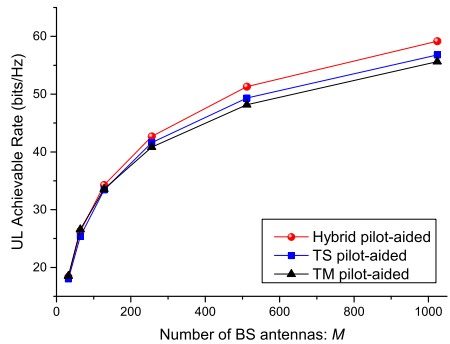}
\vspace{-5mm}
\caption{Comparison of UL achievable rate for different pilot-based schemes ($M$ = 256 and SNR = 20 dB).}
\vspace{-2mm}
\label{fig7}
\end{figure}

Finally, the performance comparison between the hybrid pilot-aided scheme and the conventional methods is demonstrated in Fig. \ref{fig7}. In these simulations, we consider the conventional schemes (that were widely considered as benchmarks in related works) employing either TM pilots \cite{YHF13} or TS pilots \cite{TVT16} (legend by ``TM pilot-aided'' and ``TS pilot-aided''), respectively, and perform data detection by using the MF detector in (\ref{detection}). For fairness of comparison, both TM pilots and TS pilots are optimized in the sense of maximizing the UL achievable rate. To elaborate a little further, TM pilots are optimally designed and reused among cells for different coherent frame-size, while TS pilots are optimized on the aspect of power allocation between pilots and data. Clearly, the hybrid pilot-aided scheme is superior to both the TM pilot-only and TS pilot-only ones \cite{YHF13} and \cite{TVT16}.

To gain an insight into the hybrid pilots, the same comparison is done in Fig. 9 for different transmission frame-size $T$. Clearly, the hybrid pilot-aided scheme performs superior to the conventional pilot-based schemes. It is also observed from Fig. 9 that, the TS pilot-based scheme \cite{TVT16} outperforms the TM pilot-only one \cite{YHF13} when frame-size $T$ is limited, whereas the TM pilot-aided method performs better than that employing only TS pilots as $T$ grows larger, i.e., when $T \gg 5KL$. This fact implies that the effectiveness for either TM pilots or TS pilots depends crucially on the frame-size $T$ in practical scenarios. In particular, we note that the gap between the UL achievable rate of hybrid pilots scheme and TM pilots one narrows down when $T \geq 10KL$, and gradually vanishes as $T$ further grows larger. This can be well explained by the results in Lemma 2 and Fig. 4 that when $T$ is large, the optimal training overhead comprises only TM pilots, i.e., $\alpha^{opt} \to 1$. Anyway, the hybrid pilot-aided scheme performs the best among all these schemes for a wide range of $T$, which demonstrates the effectiveness of the proposed design.
\begin{figure} [!t]
\centering
\includegraphics[width=3.0in]{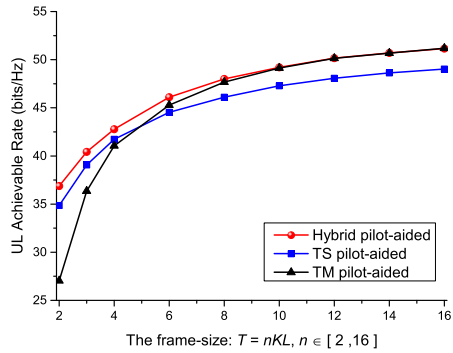}
\vspace{-5mm}
\caption{UL achievable rate v.s. the transmitted frame-size $T$ for different pilot-based schemes ($M = 256$ and SNR = 20 dB).}
\vspace{-7mm}
\label{fig8}
\end{figure}
\vspace{-3mm}

\section{Conclusion}
In this paper, we proposed a hybrid pilot-aided channel estimation scheme for multicell massive MIMO uplink, and analytically explored the impact of hybrid pilots on the UL achievable rate. Through an analysis of the relative importance, we show that higher UL rate can be achieved by employing both TM pilots and TS pilots, and additional performance improvements are gleaned by optimizing the time fraction between the two types of pilots, as well as the power and time ratio between pilots and data. Theoretical and numerical results demonstrate that the hybrid design enjoys mutual benefits between TM pilots and TS pilots, and thereby, offers a superior solution to the conventional pilot-based schemes in large MIMO systems.


\vspace{-3mm}
\begin{appendices}
\section{Derivations of UL Achievable Rate}
We discuss each term contained in (\ref{SINR}) and (\ref{SINR1}), including the power of signal, noise, self-interference, and cross-interference, respectively.

\vspace{-3mm}
\subsection{Signal and Noise Power}
From (\ref{detection1}) and (\ref{detection2}), we note that the signal and noise are identical, i.e., $\textit{\textbf{S}}(t) = \textit{\textbf{S}}^{\prime}(t)$ and $\textit{\textbf{N}}(t) = \textit{\textbf{N}}^{\prime}(t)$. Therefore, we only consider the power of $\textit{\textbf{S}}(t)$ and $\textit{\textbf{N}}(t)$, which can be given by
\begin{subequations}
\begin{align}
\label{signal_power}
E\big[|\textit{\textbf{S}}(t)|^2\big] &=E\big[\|\hat{\textbf{h}}_{1,k}\|^4\big](1-\lambda), \\
\label{noise_power}
E\big[|\textit{\textbf{N}}(t)|^2\big] &= E\big[\|\hat{\textbf{h}}_{1,k}\|^2\big]\sigma_n^2.
\end{align}
\end{subequations}
\vspace{-1mm}
We firstly evaluate the expectation $E[\| \hat{\textbf{h}}_{1,k} \|^4]$ in the above equation. Since $\{\textbf{p}_{1,k}\}$ and $\{\textbf{s}_{j,k}\}, \forall j$ are fixed, $\hat{\textbf{h}}_{1,k}$ are linear Gaussian with zero-mean and variance $\sigma_{\hat{h}_{1,k}}^2$ given by
\begin{align}
\label{sigma_h}
\sigma_{\hat{h}_{1,k}}^2 &= \beta_{1,k}+\frac{1}{(\tau \lambda)^2} \Big(\sum_{j=1}^{L}\sum_{k'=1}\beta_{j,k} |\textbf{s}_{j,k'}\textbf{p}_{1,k}^H |^2 + \|\textbf{p}_{1,k}\|^2 \sigma_n^2 \Big)\nonumber \\
&=\beta_{1,k}+\frac{\sum_{j=1}^{L}\sum_{k'=1}^{K}\alpha(1-\lambda)\beta_{j,k'}+\sigma_n^2}{\tau\lambda}.
\end{align}
Since $\|\hat{\textbf{h}}_{1,k}\|^2\sim \frac{1}{2} \sigma_{\hat{h}_{1,k}}^2 \chi^2(2M)$, where $\chi^2(2M)$ denotes the chi-square distribution with $2M$ degree of freedom, we can rewrite (\ref{signal_power}) as
\begin{align}
E\big[|\textit{\textbf{S}}(t)|^2\big] &=\sigma_{\hat{h}_1}^4 (M^2 + M)(1-\lambda).
\end{align}
Since $\tau>KL$, when $L\gg1$ and $\beta_{1,k}\gg\beta_{j,k}$, $j\neq 1$, we have
\begin{align}
\label{inque}
\beta_{1,k}\gg \frac{\sum_{j=1}^{L}\sum_{k'=1}^{K}\alpha(1-\lambda)\beta_{j,k'}+\sigma_n^2}{KL\lambda}.
\end{align}
Taking average over the distribution $\textbf{p}_{1,k}$ and $\textbf{x}_{j,k}$, and omitting some intermediate derivations, it can be easily shown from (\ref{sigma_h}) that
\begin{align}
\label{sigma_h_app}
\sigma_{\hat{h}_{1,k}}^2 = \beta_{1,k} + \mathcal{O}\Big(\frac{1}{\tau}\Big),
\end{align}
which leads to $E[\|\hat{\textbf{h}}_{1,k}\|^4] = M^2\beta_{1,k}^2 + \mathcal{O}\big(\frac{M^2}{\tau}\big) + \mathcal{O}\big(\frac{M^2}{\tau^2}\big)$. We obtain
\begin{subequations}
\begin{align}
\label{power_s}
E\big[|\textit{\textbf{S}}(t)|^2\big] &= (1-\lambda)\beta_{1,k}^2 M^2 + \mathcal{O}\Big(\frac{M^2}{\tau}\Big)  \approx (1-\lambda)\beta_{1,k}^2 M^2, \\
\label{power_n}
E\big[|\textit{\textbf{N}}(t)|^2\big] &= \sigma_n^2\beta_{1,k} M + \mathcal{O}\Big(\frac{M}{\tau}\Big) \approx \sigma_n^2\beta_{1,k} M.
\end{align}
\end{subequations}
As shown above, the power of signal and noise are of an order $\mathcal{O}(M^2)$ and $\mathcal{O}(M)$, respectively. An explanation for this behavior is the array gain that benefits from the coherent combining in a massive MIMO system of $M$ BS antennas.

\subsection{Power of Self-Interference}
We next consider the self-interference terms in (\ref{detection1}) and (\ref{detection2}),
\begin{subequations}
\begin{align}
 \label{I_1}
 \textit{\textbf{I}}_{1}(t) &= \hat{\textbf{h}}~\!\!^H_{1,k} \big(\textbf{h}_{1,k}-\hat{\textbf{h}}_{1,k}\big) x_{1,k}(t), \\
 \textit{\textbf{I}}^{\prime}_{1}(t) &= \hat{\textbf{h}}~\!\!^H_{1,k} \big(\textbf{h}_{1,k}-\hat{\textbf{h}}_{1,k}\big) s_{1,k}(t).
\end{align}
\end{subequations}
We firstly consider $\textit{\textbf{I}}_{1}(t)$, since the extension of derivation to $\textit{\textbf{I}}^{\prime}_{1}(t)$ is straightforward. For the ease of analysis, we propose decomposing $\hat{\textbf{h}}_{1,k}$ as two independent terms by pre- and post-multiplying (\ref{Hhat}) by $\hat{\textbf{h}}~\!\!_{1,k}^H$ and $\textbf{h}_{1,k}^H$, respectively. Then, we obtain the following new equation,
\begin{align}
\label{decompose}
\textbf{h}_{1,k} = \phi_{1,k} \hat{\textbf{h}}_{1,k} + \textbf{w}_{1,k},
\end{align}
where
\vspace{-2mm}
\begin{align}
\label{phi}
\phi_{1,k} = \frac{\beta_{1,k}}{\sigma_{\hat{h}_{1,k}}^2} \Big(1 + \frac{1}{\tau \lambda} \textbf{p}_{1,k} \textbf{s}_{1,k}^H \Big),
\end{align}
\vspace{-7mm}
\begin{align}
\label{w}
\textbf{w}_{1,k} = \frac{1}{\tau \lambda} \frac{\textbf{h}_{1,k}}{M \sigma_{\hat{h}_{1,k}}^2 } \Big(\sum_{k'\neq k}^{K}\big(\textbf{p}_{1,k}\textbf{s}_{1,k'}^{H}\big)\textbf{h}_{1,k'}^H+\sum_{j\neq 1}^{L}\sum_{k'=1}^{K}\big(\textbf{p}_{1,k}
\textbf{s}_{j,k'}^H\big)\textbf{h}_{j,k'}^H  + \textbf{p}_{1,k}\textbf{N}_1^H \Big) \hat{\textbf{h}}_{1,k}.
\end{align}
Due to the independency between $\phi_{1,k}\hat{\textbf{h}}_{1,k}$ and $\textbf{w}_{1,k}$, the variance of $\textbf{w}_{1,k}$, which is denoted by $\sigma_{w_{1,k}}^2$, can be given by
\vspace{-3mm}
\begin{align}
\label{sigma_w}
\sigma_{w_{1,k}}^2 = \beta_{1,k} \Big(1 - \frac{\beta_{1,k} \big|1 + \frac{1}{\tau \lambda} \textbf{p}_{1,k} \textbf{s}_{1,k}^H \big|^2}{\sigma_{\hat{h}_{1,k}}^2} \Big).
\end{align}
Substituting (\ref{phi}) and (\ref{w}) into (\ref{I_1}), we can rewrite $\textit{\textbf{I}}_{1}(t)$ as
\begin{align}
\textit{\textbf{I}}_{1}(t) = (\phi_{1,k} - 1) \| \hat{\textbf{h}}_{1,k} \|^2 x_{1,k}(t) + \hat{\textbf{h}}~\!\!_{1,k}^H \textbf{w}_{1,k} x_{1,k}(t).
\end{align}
Accordingly, the average power of $\textit{\textbf{I}}_{1}(t)$ can be given by
\begin{subequations}
\begin{align}
\label{40}
E\big[|\textit{\textbf{I}}_{1}(t)|^2\big] & = |\phi_{1,k} - 1|^2 E\big[\| \hat{\textbf{h}}_{1,k} \|^4\big] + \sigma_{w_{1,k}}^2 E\big[\| \hat{\textbf{h}}_{1,k}\|^2 \big] \\
\label{40b}
& = |\phi_{1,k} - 1|^2 \sigma_{\hat{h}_{1,k}}^4 (M^2 + M) + \sigma_{w_{1,k}}^2 \sigma_{\hat{h}_{1,k}}^2 M.
\end{align}
\end{subequations}
The equality in (\ref{40}) is due to the independency between $\hat{\textbf{h}}_{1,k}$ and $\textbf{w}_{1,k}$. By averaging $|\phi_{1,k}|^2 \sigma_{\hat{h}_{1,k}}^4$ and $\sigma_{w_{1,k}}^2 \sigma_{\hat{h}_{1,k}}^2$ in (\ref{40b}) over the distribution $\{\textbf{p}_{1,k}\}$ and $\{\textbf{s}_{j,k}\},\forall j$, we can rewrite (\ref{40b}) as

\begin{align}
\label{power_I_1}
E\big[|\textit{\textbf{I}}_{1}(t)|^2\big]
&= E\bigg[ \frac{\beta_{1,k}^2 |\textbf{p}_{1,k}\textbf{s}_{1,k}^H|^2}{(\tau\lambda)^2}  + \frac{\sum_{j=1}^{L}\sum_{k'=1}^{K}\beta_{j,k'} |\textbf{p}_{1,k} \textbf{s}_{j,k'}^H|^2 + \|\textbf{p}_{1,k}\|^4 \sigma_n^4 }{(\tau \lambda)^4}\bigg](M^2+M) \nonumber \\
&~~~ + E\bigg[\frac{\beta_{1,k}\big(\sum_{k'\neq k}^{K}|\textbf{p}_{1,k}\textbf{s}_{1,k'}^H|^2+\sum_{j \neq 1}^{L}\sum_{k'=1}^{K}\beta_{j,k'} |\textbf{p}_{1,k}\textbf{s}_{j,k'}|^2 + \|\textbf{p}_{1,k}\|^2\sigma_n^2\big)}{(\tau \lambda)^2} \bigg] M
\nonumber \\
& = \frac{(1-\lambda)\alpha}{\lambda} \cdot \frac{M^2}{\tau} \cdot \beta_{1,k}^2 + \mathcal{O}\Big(\frac{M^2}{\tau^2}\Big) + \mathcal{O}\Big(\frac{M}{\tau}\Big)\approx \frac{(1-\lambda)\alpha}{\lambda} \cdot \frac{M^2}{\tau} \cdot \beta_{1,k}^2.
\end{align}
By similarity, according to (\ref{inque}), the approximate power of $\textit{\textbf{I}}^{\prime}_{1}(t)$ for large $M$ can be derived as
\begin{align}
\label{power_I_1'}
E\big[|\textit{\textbf{I}}^{\prime}_{1}(t)|^2\big] = (1-\lambda) E\big[|\textit{\textbf{I}}_{1}(t)|^2\big] \approx \frac{\alpha(1-\lambda)^2}{\lambda} \cdot \frac{M^2}{\tau} \cdot \beta_{1,k}^2.
\end{align}
The details of (\ref{power_I_1'}) are omitted since the derivations are similar to that of $\textit{\textbf{I}}_1$. It is worth noting that the power of self-interference and signal are both in the order of $\mathcal{O}(M^2)$. That is, if $\tau$ is fixed and $\alpha \neq 0$, self-interference imposes a limit on SINR in (\ref{SINR}) even when $M \to \infty$.

\subsection{Power of Cross-Interference}
In the following, we only consider the cross-interference terms $E[|I_2(t)|^2]$ and $E[|I_3(t)|^2]$, since the extension to $I_2^{\prime}(t)$ and $I_3^{\prime}(t)$ are straightforward. From (\ref{detection1}), the power of cross-interference is given by
\vspace{-3mm}
\begin{subequations}
\begin{align}
E\big[|I_2(t)|^2\big] & = E\big[|\hat{\textbf{h}}~\!\!^{H}_{1,k} \textbf{h}_{1,k'}|^2\big],~~ k' \neq k, \\
E\big[|I_3(t)|^2\big] & = E\big[|\hat{\textbf{h}}~\!\!^{H}_{1,k} \textbf{h}_{j,k}|^2\big],~~ j \neq 1.
\end{align}
\end{subequations}

From (\ref{LS}), $\hat{\textbf{h}}_{1,k}$ and $\textbf{h}_{1,k'}$ are correlated. Again, we decompose $\textbf{h}_{1,k'}$ as
\begin{align}
\label{decompose_k'}
\textbf{h}_{1,k'} = \phi_{1,k'} \hat{\textbf{h}}_{1,k} + \textbf{w}_{1,k'},
\end{align}
\vspace{-2mm}
where
\begin{align}
\label{phi_k'}
\phi_{1,k'} = \frac{1}{\tau \lambda} \frac{\beta_{1,k'}}{\sigma_{\hat{h}_{1,k}}^2} \big(\textbf{p}_{1,k} \textbf{s}_{1,k'}^H\big),
\end{align}
\vspace{-6mm}
\begin{align}
\label{w_k'}
\textbf{w}_{1,k'} = \frac{1}{\tau\lambda \sigma_{\hat{h}_{1,k}}^2} \textbf{h}_{1,k'} \Big(\sum_{k''\neq k'}^{K}\big(\textbf{p}_{1,k} \textbf{s}_{1,k''}^{H} \big)\textbf{h}_{1,k''}^H+\sum_{j\neq 1}^L\sum_{k''=1}^{K}\big(\textbf{p}_{1,k}\textbf{s}_{j,k''}^H \big)\textbf{h}_{j,k''}^H + \textbf{p}_{1,k}\textbf{N}_j^H+\textbf{h}_{1,k} \Big)\hat{\textbf{h}}~\!\!_{1,k}^H.
\end{align}
Substituting (\ref{phi_k'}) and (\ref{w_k'}) into (\ref{decompose_k'}), we obtain
\begin{subequations}
\begin{align}
E\big[|I_2(t)|^2\big] &= E\Big[\hat{\textbf{h}}~\!\!_{1,k}^H \big(\phi_{1,k'} \hat{\textbf{h}}~\!\!^{H}_{1,k} + \textbf{w}_{1,k'} \big)\Big] \nonumber \\
\label{independent_Ik'}
& = |\phi_{1,k'}|^2 E\big[\|\hat{\textbf{h}}_{1,k}\|^4\big] + \sigma_{w_{1,k'}}^2 E\big[\|\hat{\textbf{h}}_{1,k}\|^2\big]  \\
\label{I_k'}
& = |\phi_{1,k'}|^2 \sigma_{\hat{h}_{1,k}}^4(M^2+M) + \sigma_{w_{1,k'}}^2 \sigma_{\hat{h}_{1,k}}^2 M.
\end{align}
\end{subequations}
where $\sigma_{w_{1,k'}}^2$ is the variance of $\textbf{w}_{1,k'}$ given by
\begin{align}
\label{sigma_wk'}
\sigma_{w_{1,k'}}^2 = \beta_{1,k'}\Big(1 - \frac{\beta_{1,k'}}{(\tau\lambda)^2 \sigma_{\hat{h}_{1,k}}^2} \big|\textbf{p}_{1,k} \textbf{s}_{1,k'}^H\big|^2 \Big).
\end{align}
The equalities in (\ref{independent_Ik'}) is due to the independency between $\hat{\textbf{h}}_{1,k}$ and $\textbf{w}_{1,k'}$. Similar to (\ref{power_I_1}), we obtain $E[|\phi_{1,k'}|^2 \sigma_{\hat{h}_{1,k}}^4]$ and $E[\sigma_{w_{1,k'}}^2 \sigma_{\hat{h}_{1,k}}^2]$ by averaging over the distribution $\{\textbf{p}_{1,k}\}$ and $\{\textbf{s}_{1,k}\},\forall j$, and then, we can rewrite (\ref{I_k'}) as
\begin{align}
\label{power_Ik'}
E\big[|I_2(t)|^2\big] &= E\big[|\phi_{1,k'}|^2 \sigma_{\hat{h}_{1,k}}^4\big](M^2+M) + E\big[\sigma_{w_{1,k'}}^2 \sigma_{\hat{h}_{1,k}}^2\big] M  \nonumber \\
& = E\bigg[\frac{\beta_{1,k'}^2 |\textbf{p}_{1,k} \textbf{s}_{1,k'}^H|^2}{(\tau\lambda)^2} \bigg](M^2 \!+\! M) \!+\! E\bigg[\beta_{1,k'}\Big(\beta_{1,k} \nonumber \\
&+\frac{\sum_{k''\neq k'}^{K}\beta_{1,k''}|\textbf{p}_{1,k}\textbf{s}_{j,k''}^H|^2+\sum_{j\neq 1}^{L}\sum_{k''=1}^{K}\beta_{j,k''} |\textbf{p}_{1,k}\textbf{s}_{j,k''}^H|^2 \!+\!  \|\textbf{p}_{1,k}\|^2\sigma_n^2+\beta_{1,k}}{(\tau\lambda)^2}\Big) \bigg] M  \nonumber \\
& = \frac{(1-\lambda)\alpha}{\lambda} \cdot \frac{M^2}{\tau} \cdot \beta_{1,k'}^2 + M \beta_{1,k'}\beta_{1,k} + \mathcal{O}\Big(\frac{M}{\tau}\Big) \nonumber \\
&\approx \frac{(1-\lambda)\alpha}{\lambda} \cdot \frac{M^2}{\tau} \cdot \beta_{1,k'}^2 + M \beta_{1,k'}\beta_{1,k}.
\end{align}
By analogy, the power of $I^{\prime}_2(t)$ can be derived as
\begin{align}
\label{power_Ij2}
E\big[|I^{\prime}_2(t)|^2\big] = (1-\lambda)E\big[|I_2(t)|^2\big] \approx \frac{(1-\lambda)^2\alpha}{\lambda} \cdot \frac{M^2}{\tau} \cdot \beta_{1,k'}^2 + (1-\lambda) M \beta_{1,k'}\beta_{1,k}.
\end{align}

Applying a similar procedure, we decompose $\textbf{h}_{j,k}$ as
\begin{align}
\label{decompose_j}
\textbf{h}_{j,k} = \phi_{j,k} \hat{\textbf{h}}_{1,k} + \textbf{w}_{j,k},
\end{align}
where
\begin{align}
\label{phi_j}
\phi_{j,k} = \frac{1}{\tau \lambda} \frac{\beta_{j,k}}{\sigma_{\hat{h}_{1,k}}^2} \big(\textbf{p}_{1,k} \textbf{s}_{j,k}^H\big),
\end{align}
\begin{align}
\label{w_j}
\textbf{w}_{j,k} = \frac{1}{\tau\lambda \sigma_{\hat{h}_{1,k}}^2} \textbf{h}_{j,k} \bigg(\sum_{k'\neq k}^{K}\big(\textbf{p}_{1,k} \textbf{s}_{j,k'}^{H} \big)\textbf{h}_{j,k'}^H+\sum_{i\neq j}^L\sum_{k'=1}^{K}\big(\textbf{p}_{1,k}\textbf{s}_{i,k'}^H \big)\textbf{h}_{i,k'}^H + \textbf{p}_{1,k}\textbf{N}_i^H+\textbf{h}_{1,k} \bigg)\hat{\textbf{h}}~\!\!_{1,k}^H.
\end{align}
Substituting (\ref{phi_j}) and (\ref{w_j}) into (\ref{decompose_j}), we obtain
\begin{subequations}
\begin{align}
E\big[|I_3(t)|^2\big] &= E\Big[\hat{\textbf{h}}~\!\!_{1,k}^H \big(\phi_{j,k} \hat{\textbf{h}}~\!\!^{H}_{1,k} + \textbf{w}_{j,k} \big)\Big] \nonumber \\
\label{independent_Ij}
& = |\phi_{j,k}|^2 E\big[\|\hat{\textbf{h}}_{1,k}\|^4\big] + \sigma_{w_{j,k}}^2 E\big[\|\hat{\textbf{h}}_{1,k}\|^2\big]  \\
\label{I_j}
& = |\phi_{j,k}|^2 \sigma_{\hat{h}_{1,k}}^4(M^2+M) + \sigma_{w_j}^2 \sigma_{\hat{h}_{1,k}}^2 M.
\end{align}
\end{subequations}
where $\sigma_{w_{j,k}}^2$ is the variance of $\textbf{w}_{j,k}$ given by
\begin{align}
\label{sigma_wj}
\sigma_{w_{j,k}}^2 = \beta_{j,k}\Big(1 - \frac{\beta_{j,k}}{(\tau\lambda)^2 \sigma_{\hat{h}_{1,k}}^2} \big|\textbf{p}_{1,k} \textbf{s}_{j,k}^H\big|^2 \Big).
\end{align}
The equalities in (\ref{independent_Ij}) and (\ref{sigma_wj}) are due to the independency between $\hat{\textbf{h}}_{1,k}$ and $\textbf{w}_{j,k}$. Similar to (\ref{power_I_1}), we obtain $E[|\phi_{j,k}|^2 \sigma_{\hat{h}_{1,k}}^4]$ and $E[\sigma_{w_{j,k}}^2 \sigma_{\hat{h}_{1,k}}^2]$ by averaging over the distribution $\{\textbf{p}_{1,k}\}$ and $\{\textbf{s}_{j,k}\},\forall j$, and then, we can rewrite (\ref{I_j}) as
\begin{align}
\label{power_Ij}
E\big[|I_3(t)|^2\big] &= E\bigg[\frac{\beta_{j,k}^2 |\textbf{p}_{1,k} \textbf{s}_{j,k}^H|^2}{(\tau\lambda)^2} \bigg](M^2 \!+\! M) \!+\! E\bigg[\beta_{j,k}\Big(\beta_{1,k} \nonumber \\
&+\frac{\sum_{k'\neq k}^{K}\beta_{j,k'}|\textbf{p}_{1,k}\textbf{s}_{j,k'}^H|^2+\sum_{i\neq j}^{L}\sum_{k'=1}^{K}\beta_{j,k'} |\textbf{p}_{1,k}\textbf{s}_{j,k'}^H|^2 \!+\!  \|\textbf{p}_{1,k}\|^2\sigma_n^2+\beta_{1,k}}{(\tau\lambda)^2}\Big) \bigg] M  \nonumber \\
& = \frac{(1-\lambda)\alpha}{\lambda} \cdot \frac{M^2}{\tau} \cdot \beta_{j,k}^2 + M \beta_{j,k}\beta_{1,k} + \mathcal{O}\Big(\frac{M}{\tau}\Big) \nonumber \\
&\approx \frac{(1-\lambda)\alpha}{\lambda} \cdot \frac{M^2}{\tau} \cdot \beta_{j,k}^2 + M \beta_{j,k'}\beta_{1,k}.
\end{align}
By analogy, the power of $I^{\prime}_3(t)$ can be derived as
\begin{align}
\label{power_Ij'}
E\big[|I^{\prime}_3(t)|^2\big] = (1-\lambda)E\big[|I_3(t)|^2\big] \approx \frac{(1-\lambda)^2\alpha}{\lambda} \cdot \frac{M^2}{\tau} \cdot \beta_{j,k}^2 + (1-\lambda) M \beta_{j,k}\beta_{1,k}.
\end{align}

\subsection{Correlation of Signal and Self-Interference}
From (\ref{SINR}) and (\ref{SINR1}), we observe that the signal $S(t)$ and self-interference $I_1(t)$ ($I_1^{\prime}$ in (\ref{detection2})) are correlated. This correlation may complicate the performance analysis of the data detection. To evaluate the effect of correlation, we quantify the correlation using following criteria \cite{ROSS_prob05}
\begin{align}
\label{rcorr}
\zeta=\frac{\big|E[\textit{S}^{*}(t)\textit{I}_{1}(t)]\big|^2}{E\big[|\textit{S}(t)|^2\big]E\big[|\textit{I}_1(t)|^2\big]}.
\end{align}
Since $E[|\textit{S}(t)|^2]$ and $E[|\textit{I}_1(t)|^2]$ in the denominator of (\ref{rcorr}) have been derived in (\ref{power_s}) and (\ref{power_I_1}), respectively, we consider $|E[\textit{S}^{*}(t)\textit{I}_{1}(t)]|^2$ in the numerator of (\ref{rcorr}). Following the same procedure as in deriving self-interference, we have
\begin{align}
\label{corr_S_I}
E[\textit{S}^{*}(t)\textit{I}_{1}(t)] &= E\big[\|\hat{\textbf{h}}_{1,k} \|~ \hat{\textbf{h}}~\!\!^H_{1,k} \big(\textbf{h}_{1,k} - \hat{\textbf{h}}_{1,k} \big)\big] (1-\lambda)   \nonumber \\
&= E\big[(\phi_{1,k}-1)\|\hat{\textbf{h}}_{1,k} \|^4 \big](1-\lambda) + E\big[\|\hat{\textbf{h}}_{1,k} \| \big(\hat{\textbf{h}}~\!\!^H_{1,k} \textbf{w}_{1,k}\big) \big] (1-\lambda)     \nonumber \\
&= (\phi_{1,k}-1) \sigma_{\hat{h}_{1,k}}^4 (M^2 + M) (1-\lambda).
\end{align}
Again, we derive $E[(\phi_{1,k}-1) \sigma_{\hat{h}_{1,k}}^4]$ by averaging $(\phi_{1,k}-1) \sigma_{\hat{h}_{1,k}}^4$ over the distribution $\{\textbf{p}_{1,k}\}$ and $\{\textbf{s}_{j,k}\},\forall j$
\begin{align}
\label{E_S_I}
E\big[(\phi_{1,k}-1) \sigma_{\hat{h}_{1,k}}^4\big]
 \!=\! E\Bigg[\bigg(&\frac{\textbf{p}_{1,k}\textbf{s}_{1,k}^H}{\tau\lambda}\beta_{1,k} \!-\! \sum_{j=1}^L\sum_{k'=1}^{K} \frac{|\textbf{p}_{1,k}\textbf{s}_{j,k'}^H|^2}{(\tau\lambda)^2}\beta_{j,k'} \!-\! \frac{\|\textbf{p}_{1,k}\|^2\sigma_n^2}{(\tau\lambda)^2} \bigg)  \nonumber \\
&\times \bigg(\beta_{1,k} \!+\! \sum_{j=1}^L\sum_{k'=1}^{K} \frac{|\textbf{p}_{1,k}\textbf{s}_{1,k'}^H|^2}{(\tau\lambda)^2}\beta_{j,k'} \!+\! \frac{\|\textbf{p}_{1,k}\|^2\sigma_n^2}{(\tau\lambda)^2} \bigg)\Bigg]   \nonumber \\
& = \frac{E[\textbf{p}_{1,k} \textbf{s}_{1,k}^H] \beta_{1,k}^2}{\lambda} \cdot \frac{1}{\tau} + \mathcal{O}\Big(\frac{1}{\tau^2}\Big).
\end{align}
From (\ref{inque}), and (\ref{corr_S_I})-(\ref{E_S_I}), when $M$ is large, we obtain
\begin{align}
\label{numerator}
\big|E[\textit{S}^{*}(t)\textit{I}_{1}(t)]\big|^2
& = \Big|E\big[(\phi_{1,k}-1) \sigma_{\hat{h}_{1,k}}^4\big] (M^2 + M) (1-\lambda) \Big|^2   \nonumber \\
& = \frac{(1-\lambda)^2 \beta_{1,k}^4 \big|E[\textbf{p}_{1,k}\textbf{s}_{1,k}^H]\big|^2}{\lambda^2}\cdot\frac{M^4}{\tau^2} + \mathcal{O}\Big(\frac{M^4}{\tau^4}\Big) + \mathcal{O}\Big(\frac{M^2}{\tau^2}\Big).
\end{align}
From (\ref{power_s}) and (\ref{power_I_1}), we have
\begin{align}
\label{denominator}
E\big[|\textit{S}(t)|^2\big]E\big[|\textit{I}_1(t)|^2\big] = \frac{\alpha(1-\lambda)^2 \beta_{1,k}^4}{\lambda} \cdot \frac{M^4}{\tau} + \mathcal{O}\Big(\frac{M^4}{\tau^3}\Big) +  \mathcal{O}\Big(\frac{M^3}{\tau^3}\Big).
\end{align}
Substituting (\ref{numerator}) and (\ref{denominator}) into the (\ref{rcorr}), we arrive at the following result
\begin{align}
\zeta = \frac{\frac{(1-\lambda)^2 \beta_{1,k}^4 |E[\textbf{p}_{1,k}\textbf{s}_{1,k}^H]|^2}{\lambda^2}\cdot\frac{M^4}{\tau^2} + \mathcal{O}\big(\frac{M^4}{\tau^4}\big) + \mathcal{O}\big(\frac{M^2}{\tau^2}\big)} {\frac{\alpha(1-\lambda)^2 \beta_{1,k}^4}{\lambda} \cdot \frac{M^4}{\tau} + \mathcal{O}\big(\frac{M^4}{\tau^3}\big) +  \mathcal{O}\big(\frac{M^3}{\tau^3}\big)}  & \approx \frac{|E[\textbf{p}_{1,k}\textbf{s}_{1,k}^{H}]|^2}{\alpha \lambda \cdot \tau}  \ll 1.
\end{align}
Since $\{s_{1,k}(t)\}$ and $\{p_{1,k}(t)\}$ are mutually independent, i.e. $\frac{1}{\tau}E[\textbf{p}_{1,k}\textbf{s}_{1,k}^H] \to 0$, it is reasonable to ignore the correlation between the signal and self-interference when both $M$ and $\tau$ are large.

\section{Asymptotically Optimal Solutions}
\subsection{Proof of Lemma 2}
\begin{proof}
The derivative of $\widetilde{R}_{1,k}$ in (\ref{throughputasymp_ith}) w.r.t. $\alpha$ is given by
\begin{align}
\label{derTP}
\frac{\partial \widetilde{R}_{1,k}}{\partial\alpha} &= - \Big(1-\frac{\tau}{T}+\frac{\alpha\tau}{T}\Big) \cdot \frac{1}{ln2} \cdot \frac{1}{\alpha+g}+ \frac{\tau}{T} \textmd{log}_2 \bigg(1+\frac{(1-\lambda) \beta_{1,k}^2}{\frac{(1-\lambda)\alpha}{\lambda \tau} c_1 + \frac{1}{M} \big(c_2 + \beta_{1,k} \sigma_n^2\big) } \bigg),
\end{align}
where $g= \frac{\lambda\tau(c_2 + \beta_{1,k} \sigma_n^2)}{M(1-\lambda)c_1}$. From (\ref{derTP}), we note that the linearity of $\widetilde{R}_{1,k}$ w.r.t. $\alpha$ depends on $\frac{\tau}{T}$. By deriving the second order derivative of $\widetilde{R}_{1,k}$ w.r.t. $\alpha$, we obtain
\begin{align}
\label{secondder}
\frac{\partial^2 \widetilde{R}_{1,k}}{\partial\alpha^2}&=-\frac{2\tau}{T}\cdot\frac{1}{ln2}\cdot\frac{1}{\alpha+g}+\big(1-\frac{\tau}{T}+\frac{\alpha\tau}{T}\big)\cdot\frac{1}{ln2}\cdot\frac{1}{(\alpha+g)^2} \nonumber \\
&=\Big(\frac{1}{ln2}\cdot\frac{1}{\alpha+g}\Big)\Big(-\frac{2\tau}{T}+\big(1-\frac{\tau}{T}+\frac{\alpha\tau}{T}\big)\cdot\frac{1}{(\alpha+g)}\Big).
\end{align}
Firstly, let $\frac{\partial^2 \widetilde{R}_{1,k}}{\partial\alpha^2}>0$, we easily obtain
\begin{align}
\frac{\tau}{T} < \frac{1}{1+\alpha +2g} \leq \frac{1}{2+2g}.
\end{align}
This implies that when $\frac{\tau}{T} < \frac{1}{1+\alpha +2g} \leq \frac{1}{2+2g}$, $\widetilde{R}_{1,k}$ is a convex function w.r.t. $\alpha$. By similarity, let $\frac{\partial^2 \widetilde{R}_{1,k}}{\partial\alpha^2} < 0$, we obtain
\begin{align}
\frac{\tau}{T} > \frac{1}{1+\alpha +2g} \geq \frac{1}{1+2g},
\end{align}
which means that $\widetilde{R}_{1,k}$ is a concave function w.r.t. $\alpha$ when $\frac{\tau}{T} > \frac{1}{1+\alpha +2g} \geq \frac{1}{1+2g}$.
Finally, we consider the case $\frac{\tau}{T} \in [\frac{1}{2+2g}, \frac{1}{1+2g}]$. Let $\frac{\partial^2 \widetilde{R}_{1,k}}{\partial\alpha^2} > 0$, we obtain
\begin{align}
\alpha > \frac{T}{\tau}-1-2g.
\end{align}
This implies that $\widetilde{R}_{1,k}$ is a convex function for $\alpha \in [0,\frac{T}{\tau}-1-2g)$ when $\frac{\tau}{T} \in [\frac{1}{2+2g}, \frac{1}{1+2g}]$. By analogy, let $\frac{\partial^2 \widetilde{R}_{1,k}}{\partial\alpha^2} \leq 0$ yields
\begin{align}
\alpha \leq \frac{T}{\tau}-1-2g.
\end{align}
This fact leads to $\widetilde{R}_{1,k}$ a concave function for $\alpha \in [\frac{T}{\tau}-1-2g,1]$ when $\frac{\tau}{T} \in [\frac{1}{2+2g}, \frac{1}{1+2g}]$.

This concludes the proof.
\end{proof}

\subsection{Proof of Lemma 3}
\begin{proof}
Taking the first derivative of $\widetilde{R}_{1,k}$ w.r.t. $\lambda$, and omitting some intermediate derivations, we have
\begin{align}
\label{throughputderi}
\frac{\partial \widetilde{R}_{1,k}}{\partial\lambda} = \frac{\alpha\tau}{T}\cdot\frac{1}{ln2}\cdot\frac{-(\frac{\alpha}{\lambda \tau} + \frac{f}{(1-\lambda)})+\frac{\alpha}{\lambda^2 \tau}}{\frac{(1-\lambda)\alpha}{\lambda \tau}+ f}+\Big(1-\frac{\tau}{T}\Big)\cdot\frac{1}{ln2}\cdot \frac{\frac{\alpha}{\lambda^2 \tau}}{\frac{(1-\lambda)\alpha}{\lambda \tau}+ f},
\end{align}
where $f=\frac{c_2+ \beta_{1,k} \sigma_n^2}{Mc_1}$. Simplifying some intermediate derivations and setting (\ref{throughputderi}) to zero, we arrive at the following second-order polynomial of $\lambda$,
\begin{align}
\label{lambdaopteq}
\big(\alpha\tau-\tau^2f\big)\lambda^2-\big(T-\tau+2\alpha\tau\big)\lambda+\big(T-\tau+\alpha\tau\big)=0.
\end{align}
It is easily to show that $\widetilde{R}_{1,k}(\lambda)$ is a concave function. Hence, the global maximum can be obtained by solving the above formula. The optimal $\lambda$ that maximizes $\widetilde{R}_{1,k}$ can be given by
\begin{align}
\label{lambdaopt1}
\lambda^{opt}=\frac{(T-\tau)+2\alpha\tau-\sqrt{(T-\tau)^2+4\tau^2f\big((T-\tau)+\alpha\tau\big)}}{2(\alpha\tau-\tau^2f)}.
\end{align}
This concludes the proof.
\end{proof}

\subsection{Proof of Lemma 4}
\begin{proof}
The first derivative of $\widetilde{R}_{1,k}$ w.r.t. $\tau$ is given by
\begin{align}
\label{throughputderttau}
\frac{\partial \widetilde{R}_{1,k}}{\partial\tau} &=
\frac{\alpha}{T}\textmd{log}_2 \bigg(1+\frac{(1-\lambda) \beta_{1,k}^2}{\frac{(1-\lambda)\alpha}{\lambda \tau}c_1 +\frac{c_2 + \beta_{1,k} \sigma_n^2}{M} } \bigg) -\frac{1}{T}\textmd{log}_2 \bigg( 1 + \frac{\beta_{1,k}^2} {\frac{(1-\lambda)\alpha}{\lambda \tau}c_1 + \frac{c_2 + \beta_{1,k} \sigma_n^2}{M}} \bigg) \nonumber \\
&+\Big(\frac{1}{\tau}-\frac{1-\alpha}{T}\Big)\cdot\frac{1}{ln2}\cdot\frac{\frac{(1-\lambda)\alpha}{\lambda \tau} c_1}{\frac{(1-\lambda)\alpha}{\lambda \tau}c_1 + \frac{c_2+ \beta_{1,k} \sigma_n^2}{M}}.
\end{align}
Let $h = \frac{\lambda}{(1-\lambda)\alpha M}\frac{c_2 + \beta_{1,k} \sigma_n^2}{c_1}$, the second order derivative of $\widetilde{R}_{1,k}$ w.r.t. $\tau$ can be given by
\begin{align}
\label{throughputderttausec}
\frac{\partial^2 \widetilde{R}_{1,k}}{\partial\tau^2} = -\frac{1}{ln2} \bigg(\underbrace{\frac{1-\alpha}{T} \cdot\frac{1}{\tau(1+h\tau)}}_{\geq 0} + \underbrace{\frac{T-(1-\alpha)\tau}{T\tau} \cdot\frac{h}{(1+h\tau)^2}}_{\geq 0} + \underbrace{\frac{1}{(1+h\tau)\tau^2}}_{> 0} \bigg).
\end{align}
From the above, the second order derivative on $\tau$ is strictly negative, implying that $\widetilde{R}_{1,k}$ is a concave function respect to $\tau$. From (\ref{throughputderttau}), we have the following observations:
\begin{itemize}
  \item $C1$: When $\tau = KL, \frac{\partial \widetilde{R}_{1,k}}{\partial\tau} < 0$, which implies that $\widetilde{R}_{1,k}$ is a monotonically decreasing function with $\tau^{opt} = KL$.
  \item $C2$: When $\frac{\partial \widetilde{R}_{1,k}}{\partial\tau}|_{\tau = T} > 0$, $\widetilde{R}_{1,k}$ is a monotonically increasing function with $\tau^{opt} = T$.
  \item $C3$: When $\frac{\partial \widetilde{R}_{1,k}}{\partial\tau}|_{\tau < T} \geq 0$, $\widetilde{R}_{1,k}$ is a concave function with $\tau^{opt} \in (KL,T)$.
\end{itemize}

Next, we focus on the necessary and sufficient conditions of the above cases. From (\ref{throughputderttau}), we note that $\frac{\partial \widetilde{R}_{1,k}}{\partial\tau}$ depends on $\alpha$, and thus, take the first order derivative of (\ref{throughputderttau}) w.r.t. $\alpha$ as
\begin{align}
\label{throughputderttaualpha}
\frac{\partial^2 \widetilde{R}_{1,k}}{\partial\tau \partial\alpha}
&= \frac{1}{T}\textmd{log}_2\bigg(1+\frac{(1-\lambda) \beta_{1,k}^2}{\frac{(1-\lambda)\alpha}{\lambda \tau}c_1 + \frac{1}{M} \big(c_2+ \beta_{1,k} \sigma_n^2\big) } \bigg)  \nonumber \\
& + \frac{1}{ln2} \cdot \frac{1}{(\alpha+h)T} \bigg( (1-\alpha) + \frac{\alpha^2+2\alpha h}{\alpha+h} + \frac{(T-\tau)h}{(\alpha+h)\tau} \bigg) > 0.
\end{align}
As a consequence, $\frac{\partial \widetilde{R}_{1,k}}{\partial\tau}$ is an increasing function respect to $\alpha$. Thus, $\alpha = 1$ maximizes $\frac{\partial \widetilde{R}_{1,k}}{\partial\tau}$, which can be denoted by $\textrm{max}\{\frac{\partial \widetilde{R}_{1,k}}{\partial\tau}\}{|}_{\alpha = 1}$. Provided that $\textrm{max}\{\frac{\partial \widetilde{R}_{1,k}}{\partial\tau}\}{|}_{\alpha = 1} < 0$, $\widetilde{R}_{1,k}$ is a monotonically decreasing function w.r.t. $\tau, \tau \in [KL,T]$, which implies that $\tau^{opt} = KL$ is the optimal time-allocation to the training overhead that maximizes the UL rate $\widetilde{R}_{1,k}$. This result is identical to that of $(C1)$ with the corresponding necessary and sufficient condition as
\begin{align}
\label{conKL}
\textrm{max}\{\widetilde{R}_{1,k} \} \big{|}_{{\alpha = 1}\atop{\tau = KL}} = \frac{1}{T}\textmd{log}_2(1-\lambda) + \frac{1}{ln2} \cdot \frac{1}{(1+KL h)KL} < 0,
\end{align}
which is equivalent to
\begin{align}
\label{lambda_cond}
1 > \lambda > 1-\frac{1}{2^{\frac{T}{(1+KL h)\cdot KL ln2}}}.
\end{align}
Using the same argument, the necessary and sufficient condition of $(C2)$ is given by
\begin{align}
\label{conT}
\textrm{max}\{\widetilde{R}_{1,k} \} \big{|}_{{\alpha = 1}\atop{\tau = KL}} = \frac{1}{T}\textmd{log}_2(1-\lambda) + \frac{1}{ln2}\cdot\frac{1}{(1+Th)KL} > 0,
\end{align}
which is equivalent to
\begin{align}
\label{lambda_cond1}
0 < \lambda < 1-\frac{1}{2^{\frac{T}{(1+KL h)\cdot KL ln2}}}.
\end{align}
\end{proof}
This concludes the proof.

\end{appendices}

\ifCLASSOPTIONcaptionsoff
  \newpage
\fi

\end{document}